\documentclass[fleqn,usenatbib]{mnras}

\usepackage{newtxtext,newtxmath}

\usepackage[T1]{fontenc}

\DeclareRobustCommand{\VAN}[3]{#2}
\let\VANthebibliography\thebibliography
\def\thebibliography{\DeclareRobustCommand{\VAN}[3]{##3}\VANthebibliography}

\usepackage{graphicx}	
\usepackage{amsmath}	

\usepackage{xcolor} 
\newcommand{\ditto}{$- \prime \prime -$}
\newcommand{\dg}{^{\circ}}

\newcommand{\appropto}{\mathrel{\vcenter{
  \offinterlineskip\halign{\hfil$##$\cr
    \propto\cr\noalign{\kern2pt}\sim\cr\noalign{\kern-2pt}}}}}

\title[Repeated gamma-ray flares in blazars]{Repeated patterns of gamma-ray flares suggest structured jets of blazars as likely neutrino sources}

\author[P. Novikova et al.]{
Polina Novikova,$^{1}$\thanks{E-mail: p.a.novikova@mail.ru} Ekaterina Shishkina,$^{1}$ and Dmitry Blinov$^{2,3}$
\\
$^{1}$St. Petersburg State University, Universitetsky pr. 28, Petrodvoretz, 198504 St. Petersburg, Russia\\
$^{2}$Institute of Astrophysics, Foundation for Research and Technology - Hellas, Voutes, 7110 Heraklion, Greece\\
$^{3}$Department of Physics, University of Crete, 71003, Heraklion, Greece\\
}

\date{Accepted: 2023 September 6; Revised: 2023 September 6; Received: 2023 April 25}

\pubyear{2023}

\begin{document}
\label{firstpage}
\pagerange{\pageref{firstpage}--\pageref{lastpage}}
\maketitle

\begin{abstract}
{\em Fermi}-LAT observations provide continuous and regularly-sampled measurements of gamma-ray photon flux for hundreds of blazars. Many of these light curves, spanning almost 15 years, have been thoroughly examined for periodicity in multiple studies. However, the possibility that blazars may exhibit irregularly repeating flaring patterns in their gamma-ray light curves has not been systematically explored. In this study, we aim to find repeating episodes of flaring activity in the 100 brightest blazars using Fermi-LAT light curves with various integration times. We use a Bayesian Blocks representation to convert the time series into strings of symbols and search for repeating sub-strings using a fuzzy search algorithm. As a result, we identify 27 repeated episodes in the gamma-ray light curves of 10 blazars. We find that the patterns are most likely produced in structured jets composed of a fast spine and a slower sheath. When individual emission features propagate in the spine, they scatter seed photons produced in the non-uniform sheath through the inverse Compton mechanism, resulting in a set of gamma-ray flares with a similar profile every such passage. Additionally, we explore the theoretically-predicted possibility that the spine-sheath structure facilitates the production of high-energy neutrinos in blazar jets. Using the catalogue of track-like events detected by the IceCube neutrino telescope, we find evidence supporting this hypothesis at a $2.8\sigma$ significance level.
\end{abstract}




\section{Introduction} \label{sec:intro}

Since its launch in 2008, the Large Area Telescope \citep[LAT,][]{Atwood2009}, onboard the {\em Fermi} Gamma-ray Space Telescope, has scanned the entire sky at energies of 20 MeV to 300 GeV every three hours.
The latest version of the Fermi-LAT source catalog, 4FGL-DR3 \citep{Abdollahi2020,Abdollahi2022}, includes 6658 sources, of which $\sim$3700 are confidently identified as blazars. Blazars are a type of Active Galactic Nuclei (AGN) with a relativistic jet pointing towards our line of sight. The high energy emission of the jet is thought to be produced by inverse-Compton (IC) scattering of low-energy photons by energetic electrons, which are also responsible for the synchrotron radiation received from these objects at low frequencies. The seed photons in this process can be the same synchrotron photons produced internally in the jet. This is referred to as the "synchrotron self-Compton" \citep[SSC, e.g.,][]{Konigl1981,Kirk1998} mechanism. Alternatively, the seed photons undergoing the IC scattering can come from regions outside the jet, such as the broad-line region (BLR), the accretion disk, the dust torus, or other sources. In this case, it is called the "external Compton" \citep[EC, e.g.,][]{Dermer1992,Sikora2002} mechanism. An alternative hadronic scenario, explaining the gamma-ray emission as  synchrotron emission from protons or from secondary decay products of charged pions, is also considered in the literature \citep{Bottcher2013}.

Gamma-ray light curves of blazars are typically well described by noise processes of different types \citep{Abdo2010}. Nevertheless, some blazars exhibit periodic or quasi-periodic variability in their low frequency emission \citep{Valtaoja2000} as well as in gamma-rays \citep{Ackermann2015}. There are studies that attempt to search for such periodic behaviour using {\em Fermi}-LAT data in a systematic manner \citep{Penil2020,Penil2022}. 

There is another type of a regular behaviour in the gamma-ray light curves that is poorly investigated and that we refer to as "repeated patterns" of flares. There are two cases of such events discussed in the literature. \cite{Jorstad2013} reported about three repetitions of a series of three flares in the blazar 3C~454.3. The triplet of flares had the same duration, similar profile in all three repetitions, while the lags between individual flares were constant. Two out of three recurrences of the pattern coincided with a passage of a new knot through the radio core at 43 GHz. Therefore, it was concluded by the authors that the emission zone responsible for the flaring pattern is located in the radio core $\sim$15 - 20 pc downstream from the black hole. The three flares of the pattern in this case could be associated with a passage of a moving emission feature (shock) through a system of conical recollimation shocks within the core. In another blazar 3C~279 four repetitions of a complex pattern of flares were detected in coincidence with optical polarization plane rotations \citep{Blinov2021}. Three out of four events were followed by the ejections of new radio knots at 43 GHz. The four patterns in this case had very different durations in the observer's frame. The shortest event was $\sim 4.5$ faster than the longest. Based on the EC nature of the gamma-ray flares of the pattern and the derived size of the emission zone, \cite{Blinov2021} concluded that the shock-shock interaction is not a viable model in the case of 3C~279. They suggested that this series of repeated patterns is the predicted manifestation of the model discussed by \cite{Marscher2010}. In this model an emission feature travelling within the fast spine of the jet passes through a system of qasi-stationary ring-like condensations located in the slow sheath of the jet. Photon field created by these regions is highly relativistically boosted in the reference frame of the moving component and vice versa \citep{Ghisellini2005,MacDonald2015}. Therefore, every such passage is accompanied by a period of efficient IC scattering of these photons. In the case when the system of such rings is persistent, one can observe very similarly looking patterns of EC gamma-ray flares separated by years apart.

\cite{Hervet2019} searched for a repeated pattern of X-ray flares of Mkn 421 using a statistical approach. They demonstrated evidence at a $3 \sigma$ level that such pattern is indeed present in the light curve. Their interpretation of this finding is similar to the one of \cite{Jorstad2013} -- individual flares of the pattern are produced when a moving emission feature passes through a system of stationary recollimation shocks in the jet. There is no systematic search for repeated patterns in the gamma-ray band reported in the literature.

In recent years, there has been growing evidence of a possible association between blazars and astrophysical neutrinos \citep{Plavin2022}. Such an association was theoretically suggested at the dawn of neutrino astronomy \citep[e.g.,][]{Berezinsky1981}. Since 2013, when the IceCube observatory started regular observations of neutrinos with energies ranging from TeV to PeV \citep{Aartsen2013}, observational tests of the link between these particles and flaring blazars have become possible. The first evidence of a blazar-neutrino association was reported in 2018 when IceCube-170922A was found to be positionally consistent with TXS 0506+056, which was flaring at the moment of the particle detection \citep{Aartsen2018b}. Subsequent analyses of archival data from IceCube and other neutrino observatories have revealed additional candidate neutrino events that may be associated with other blazars. However, the statistical significance of these individual associations is rather low.

This paper describes an algorithm that can identify repeated patterns in well-sampled light curves. We applied this algorithm to the gamma-ray light curves of the 100 brightest blazars, as provided by {\em Fermi}-LAT, and report our findings. We investigate the hypothesis that the characteristic structure of the jet responsible for the observed patterns may also efficiently produce high-energy neutrinos. We present statistical evidence supporting this hypothesis.

In this work, we adopt the formal definition of a flare proposed by \cite{Nalewajko2013}. We define a flare as "a contiguous period of time, associated with a given flux peak, during which the flux exceeds half of the peak value. Furthermore, this lower threshold is achieved precisely twice - at the onset and conclusion of the flare."

\section{Sample selection and data reduction}
\subsection{Sample and gamma-ray data} \label{subsec:gamma}
We selected all sources from the 4FGL 8-year Source Catalog\footnote{\url{https://fermi.gsfc.nasa.gov/ssc/data/access/lat/8yr_catalog/gll_psc_v20.fit}} \citep{Abdollahi2020} that belong to the following categories: "BCU" and "bcu" - active galaxies of uncertain type; "FSRQ" and "fsrq" - the flat spectrum radio quasars; "bll" and "BLL" - BL Lacertae objects. Capital letters in the designations indicate firm identifications, while lowercase letters indicate associations. Then we sorted this sample according to the "Flux1000" column value and selected the top 100 sources with the highest photon flux integrated in the energy range 1 - 100 GeV.
For these 100 sources we used data available in the Fermi LAT Light Curve Repository \cite[LCR,][]{Abdollahi2023} with 3- and 7-day integration time.

In the same way, we also selected 30 "brightest" sources for which we analysed {\em Fermi}-LAT data with 1 d integration time. The data were processed in the energy range $100\, {\rm MeV} \le E \le 300\, {\rm GeV}$ using the unbinned likelihood analysis of the standard {\em Fermi} analysis software package Fermitools (v. 1.2.23) distributed under Conda. We used the instrument response function $P8R3\_SOURCE\_V3$. Source class photons (evclass=128 and evtype=3) were selected within a $15\dg$ region of interest (ROI) centered on a blazar. The Earth limb background was excluded by limiting the satellite zenith angle ($< 90\dg$). The Galactic interstellar emission was accounted for using the $gll\_iem\_v07$ spatial model. The extragalactic diffuse and residual instrumental backgrounds were included in the fit as an isotropic spectral template $iso\_P8R3\_SOURCE\_V3\_v1.txt$. The background models were created including all sources from the 4FGL within $15\dg$ of the blazar. For sources beyond $10\dg$ from the blazar, photon fluxes were fixed to their values reported in 4FGL. Similarly, for all targets in the ROI, their spectral shapes were fixed. In order to make the 1 d binned light curves consistent with 3 and 7 d data from the LCR, we set the test statistic ${\rm TS} = 10$ as the detection threshold. This corresponds to approximately a $3\sigma$ detection level \citep{Nolan2012}. The systematic uncertainties in the effective LAT area do not exceed 10 per cent in the energy range we use \citep{Ackermann2012}. Since our analysis is based on relative flux variations and because the statistical errors dominate at the short time scales analysed in this paper, we did not take into account the systematic uncertainties.

\subsection{Neutrino data} \label{subsec:neutr}

IceCube has the ability to identify high-energy neutrino events in two distinct forms: cascades and tracks. Among these, the track-like events are of utmost significance when it comes to linking high-energy neutrinos with astrophysical sources, since they provide the smallest uncertainties on the arrival direction, typically around $1\dg$ \citep{Aartsen2017a}. The probability of a neutrino originating from an astrophysical source strongly depends on its energy. At lower energies, the majority of detected neutrinos are generated in the atmosphere. For events with energy above the 200 TeV threshold, the probability of their astrophysical origin becomes $>50\%$ \citep{IceCube2019}. Therefore, to create a uniform sample of neutrino events suitable for statistical analysis, we followed the selection criteria from \cite{Plavin2020} and \cite{Hovatta2021}. We selected neutrinos with $E \ge 200$ TeV from the recently published list of track-like events IceCube Event Catalog of Alert Tracks \citep[ICECAT-1,][]{Abbasi2023}, and introduced a cut in positional accuracy so that the 90\% containment area $\Omega_{90} \le 10 {\,\rm deg}^2$. This resulted in a list of 54 neutrinos between 2011 September 2 and 2020 December 9, which is available as supplementary material to this paper.

\subsection{Optical data} \label{subsec:opt}

We collected multi-band optical data for our sample sources from publicly available repositories. Specifically, we used V and SDSS-g bands photometry from the All-Sky Automated Survey for Supernovae \citep[ASAS-SN,][]{Shapee2014}, SDSS-g, SDSS-r, and SDSS-i bands photometry from the Panoramic Survey Telescope and Rapid Response System \citep[Pan-STARRS1,][]{Chambers2016}) and  The Zwicky Transient Facility \citep[ZTF,][]{Masci2019}, as well as the R-band data from the Small and Moderate Aperture Research Telescope System \citep[SMARTS,][]{Bonning2012}. Additionally, we incorporated the "white light" observations of the Katzman Automatic Imaging Telescope \citep[KAIT,][]{Li2003}, which has an effective wavelength close to the R-band.


\section{Pattern search method} \label{sec:meth}

Our objective is to identify segments of the gamma-ray light curves spanning $\sim$15-years that closely resemble each other and have durations of tens to hundreds of days. Depending on the integration time, each light curve of the sample sources consists of $10^3$ to $5\times10^3$ measurements. A visual search for repeated patterns would be very time-consuming. In order to avoid this, we automated the search process. We investigated several possibilities, such as dividing a curve into approximately equal intervals and comparing them using various similarity measures, including the Euclidean distance and TWED \citep[Time-Warp Edit Distance,][]{Lin2012}, as well as calculating the correlation coefficient between these segments. However, due to the noisy data, the length of the light curves, the unknown duration of repeated patterns, and the possibility of time scale stretching or squeezing (which is discussed later), we ended up with many thousands of candidate patterns that required visual verification. We found that a method representing the light curves as strings of characters, similar to Symbolic Aggregate Approximation \citep{Lin2007}, followed by a fuzzy search for repeated sub-strings, is much more efficient for our task. We describe this algorithm in more detail below.

\begin{figure*}
	\includegraphics[width=0.99\textwidth]{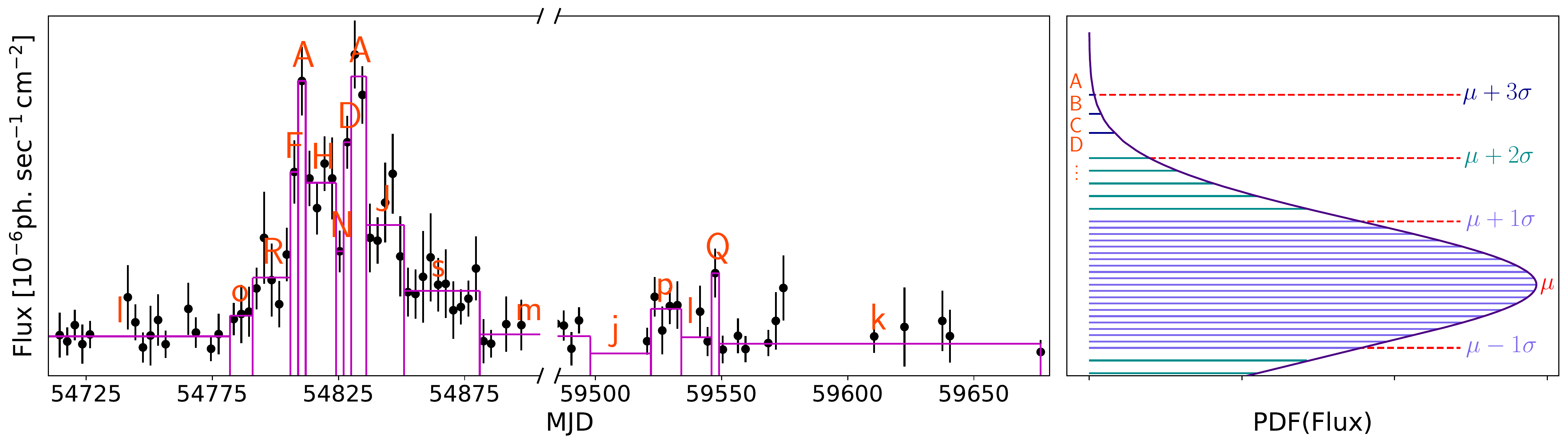}
    \caption{An example of a light curve with the Bayesian blocks approximation and the transformation to a string of symbols.}
    \label{fig:bbexpl}
\end{figure*}

To start, we represented each light curve as a step function using the Bayesian Block algorithm \citep{Scargle2013}. This algorithm approximates the data as a piecewise constant representation and identifies the optimal segmentation of the observational data. We used the $bayesian\_blocks$ function in the {\em stats} package of Astropy\footnote{\url{https://docs.astropy.org/en/stable/api/astropy.stats.bayesian_blocks.html}}. Then, for each Bayesian block, we assigned a symbol based on the flux value (block height). We used two different algorithms for this task. In the first case, using the original light curve, we computed the mean value $\mu$ and standard deviation $\sigma$ of the photon flux. We then divided the interval [$\mu$-$\sigma$, $\mu$+$\sigma$] into 20 equal parts, intervals [$\mu$-2$\sigma$, $\mu$-$\sigma$) and ($\mu$+$\sigma$, $\mu$+2$\sigma$] into five parts each, and intervals [$\mu$-3$\sigma$, $\mu$-2$\sigma$) and ($\mu$+2$\sigma$, $\mu$+3$\sigma$] into three parts each. This resulted in 38 intervals that cover the entire range of possible photon flux values and provide denser binning in the interval where most measurements concentrate (see Fig.\ref{fig:bbexpl}). We assigned each of the 38 bins a capital or lowercase letter in the range from "a" to "s". In the second case, we divided the interval between the maximum and minimum photon flux of the light curve into 38 equal bins and assigned each of them a unique symbol in the same range. After this procedure, each light curve was transformed into two different strings of letters.

The repeated patterns we search for are associated with individual emission features that can propagate in the jet with different Doppler factors, $\delta$ \citep{Blinov2021}. If two emission features with $\delta_1$ and $\delta_2$ produce repeated patterns, their timescales will be related as $\Delta t_1 / \Delta t_2 = \delta_2 / \delta_1$. Moreover, the flux of flares in the patterns is altered by changes in $\delta$ in a way that depends on the emission mechanism. If the emission is of synchrotron self-Compton nature, then one would expect $F^{SSC}_1/F^{SSC}_2 = (\delta_1/\delta_2)^{3+\alpha}$ \citep{Dermer1995}. In the case when the gamma-ray emission is produced by the external Compton mechanism, this dependence would be $F^{EC}_1/F^{EC}_2 = (\delta_1/\delta_2)^{4+2\alpha}$ \citep{Ghisellini2005}. In fact, the latter formula is more complex when the seed photons for the EC process are produced in a jet sheath that has non-zero $\delta$. However, we consider only the stationary sheath case for simplicity.

Our pattern search method is not sensitive to timescale stretching or compressing since we loose all temporal information converting light curves intro strings of symbols. In order to account for possible changes in the amplitude of patterns, for each light curve, we generated a grid of its transformed versions considering both SSC and EC cases. We allowed $\delta$ to vary by a factor in the range from 0.5 to 2.0 with a step 0.1. This range was chosen based on the results of \cite{Jorstad2005}, where the ratio of maximum to minimum $\delta$ of detected knots in individual bright gamma-ray blazars is $\le4$. The spectral index $\alpha$ was permitted to vary between -2.5 and 2.5, with a step size of 0.2, effectively encompassing the entire range of potential values \citep{Fossati1998}. As a result, for both emission mechanisms, we produced 390 variations of the light curve. Then we converted all these curves into strings of symbols using the same transformation as for the original (observed) light curve.

The resulting strings were divided into equal sub-strings (we used 50 - 60 characters, depending on the light curve integration time). In order to find whether these sub-strings repeated elsewhere in the entire "observed" string, we used the {\em find\_near\_matches} function from the {\em fuzzysearch} Python package\footnote{\url{https://github.com/taleinat/fuzzysearch}}. Fuzzy String Search is a method for closely matching stings instead of exactly. When calling the function, we specified the maximum of the Levenshtein distance (a metric that measures the difference between two sequences of symbols) equal to $30\%$ of the length of the sub-string under study. This means that matched sub-strings could differ by no more than $30\%$ of their length. As a result, {\em find\_near\_matches} returns the position in the string that matched the searched sub-string, if such a match was found.

We transformed the indices of the found repeated sub-strings back into modified Julian dates (MJD) corresponding to the original light curves, and then plotted these intervals for visual inspection. For each blazar with a found repeated pattern, we visually scanned the entire light curve to ensure that the automated algorithm did not miss any other repetitions of the same pattern. Finally, for all detected patterns, we fine-tuned the value of $\delta_i/\delta_1$ by finding the best agreement between the corresponding pieces of the light curve through stretching or compressing them. To achieve this, we minimized the TWED between the matched intervals of the light curve as a function of the multiplication factor and the shift of the time axis for one of the intervals. The uncertainties of $\delta_i/\delta_1$ were estimated using the bootstrap approach. For each light curve, we generated 500 realizations where each point was re-sampled from a normal distribution with a mean and standard deviation equal to the original point's photon flux and its uncertainty, respectively. For every generated light curve, we repeated the TWED minimization procedure using the PySwarms\footnote{\url{https://pyswarms.readthedocs.io/en/latest/intro.html}} Python implementation of the Particle Swarm Optimization algorithm. Finally, we calculated the standard deviation of the obtained optimal $\delta_i/\delta_1$ values for the set of generated curves. We consider this value as the uncertainty for the corresponding value defined for the actual light curve.

\section{Found patterns} \label{sec:res}

\begin{figure*}
	\includegraphics[width=0.8\textwidth]{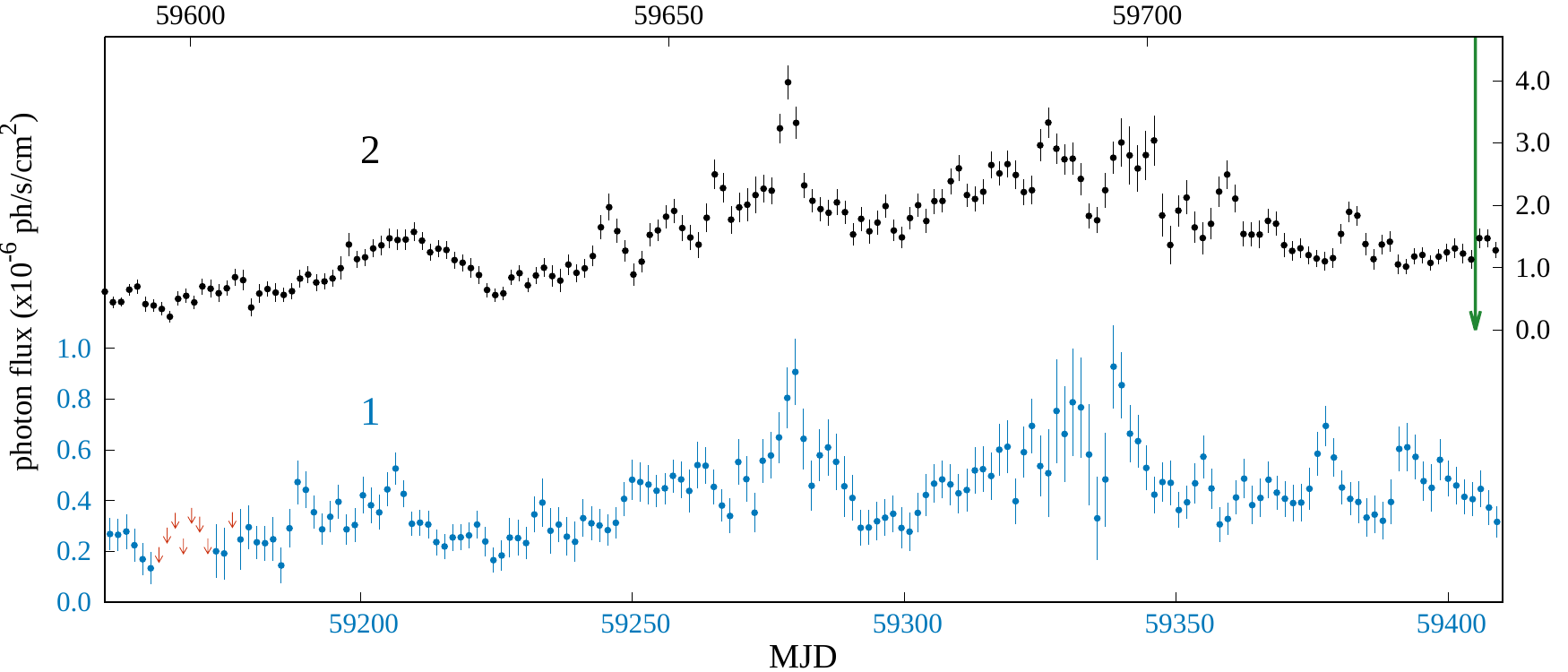}
    \caption{Repeated pattern of gamma-ray flares in 4FGL J0108.6+0134. The time scale of the curve 2 is stretched by a factor of 1.76. The integration time is 3 and 1.7 d for curves 1 and 2, respectively. The time difference between consecutive points is 1/2 of the bin duration. The green arrow indicates arrival time of GVD210710CA neutrino.}
    \label{fig:0108p1}
\end{figure*}

\begin{figure*}
	\includegraphics[width=0.8\textwidth]{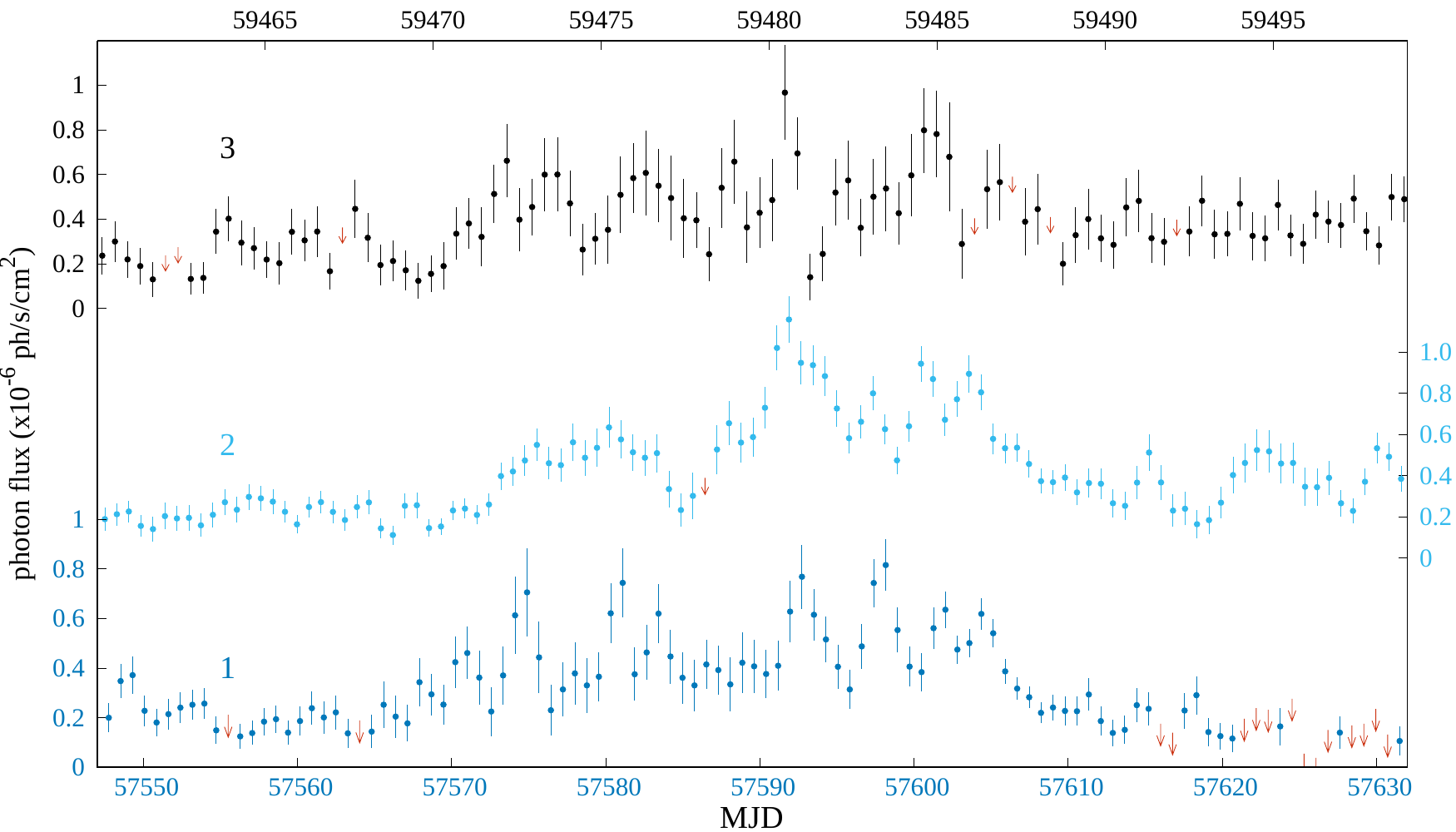}
    \caption{Repeated pattern of gamma-ray flares in 4FGL J0457.0$-$2324. The time scales of the curves 2 and 3 are multiplied by factors 0.75 and 2.2. The integration time is 1.55, 2.07 and 0.75 d for curves 1, 2 and 3, respectively. The time difference between consecutive points is 1/2 of the bin duration.}
    \label{fig:0457p1}
\end{figure*}

\begin{figure*}
	\includegraphics[width=0.8\textwidth]{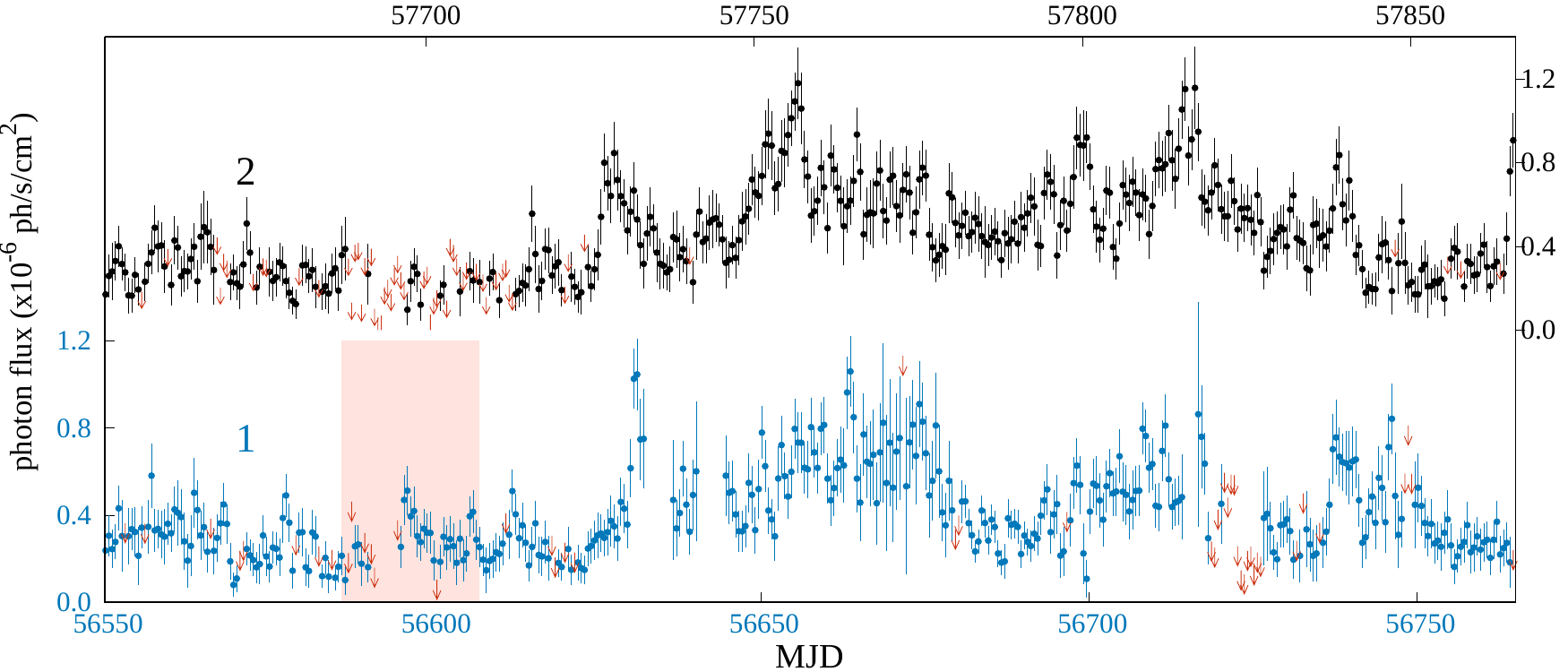}
    \caption{Repeated pattern of gamma-ray flares in 4FGL J1048.4+7143. Both curves are in the original time scale. The integration time for both curves is 1 d with 1/2 days between consecutive points. The pink region marks the period when the optical polarization plane of the blazar was rotating \citep{Blinov2015}.}
    \label{fig:1048p1}
\end{figure*}

\begin{figure*}
	\includegraphics[width=0.8\textwidth]{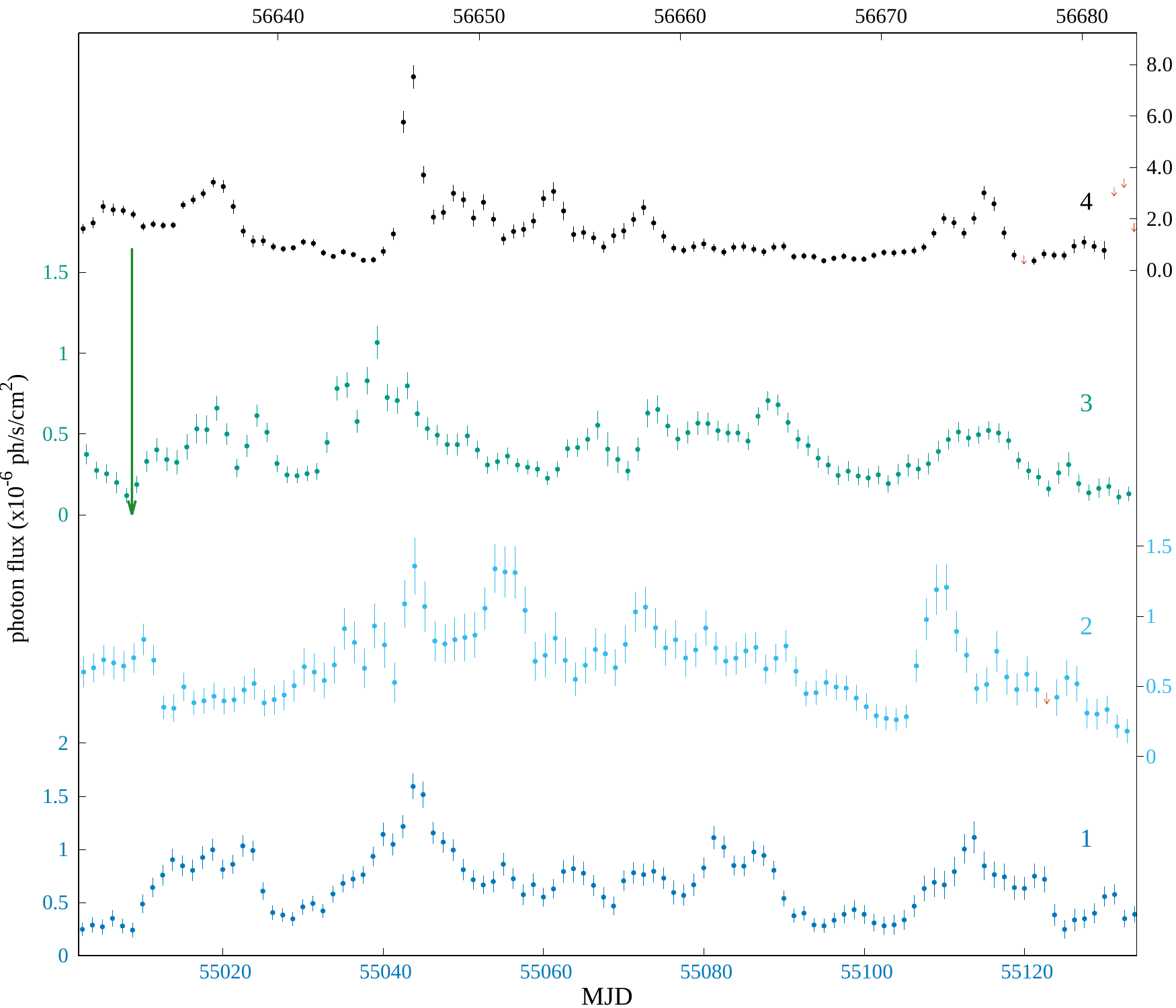}
    \caption{Repeated pattern of gamma-ray flares in 4FGL J1256.1$-$0547. The time scales of the curves 2,3 and 4 are multiplied by factors 1.582, 0.56 and 2.51. The integration time for the curves is $t_{\rm int}=$ 2.5, 1.58, 4.46 and 1.0 d with $t_{\rm int}/2$ d between consecutive points. The green arrow indicates the arrival time of the IceCube EHEA event at MJD=57291.9.}
    \label{fig:1256p1}
\end{figure*}

\begin{figure*}
	\includegraphics[width=0.8\textwidth]{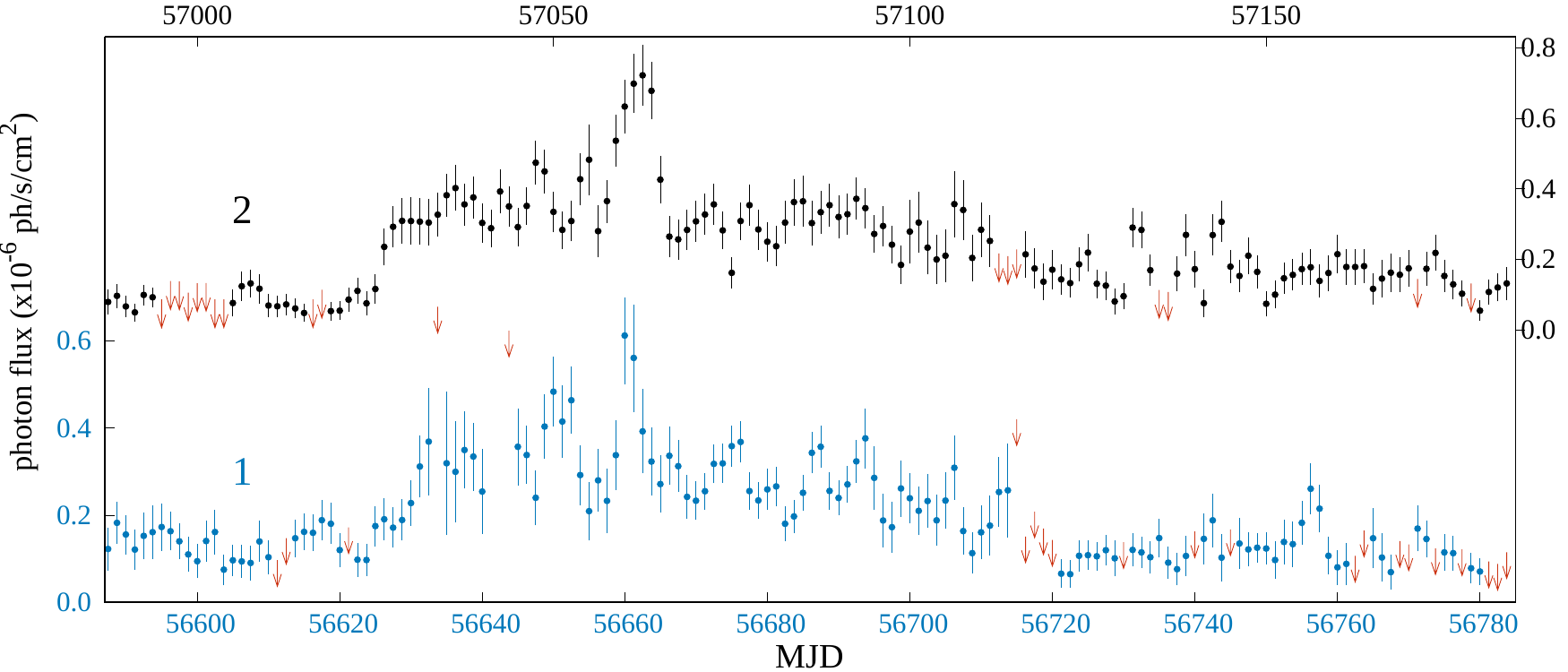}
    \caption{Repeated pattern of gamma-ray flares in 4FGL J1345.5+4453. Both curves are in the original time scale. The integration time for both curves is $t_{\rm int}=2.5$ d with $t_{\rm int}/2$ d between consecutive points.}
    \label{fig:1345p1}
\end{figure*}

\begin{figure*}
	\includegraphics[width=0.8\textwidth]{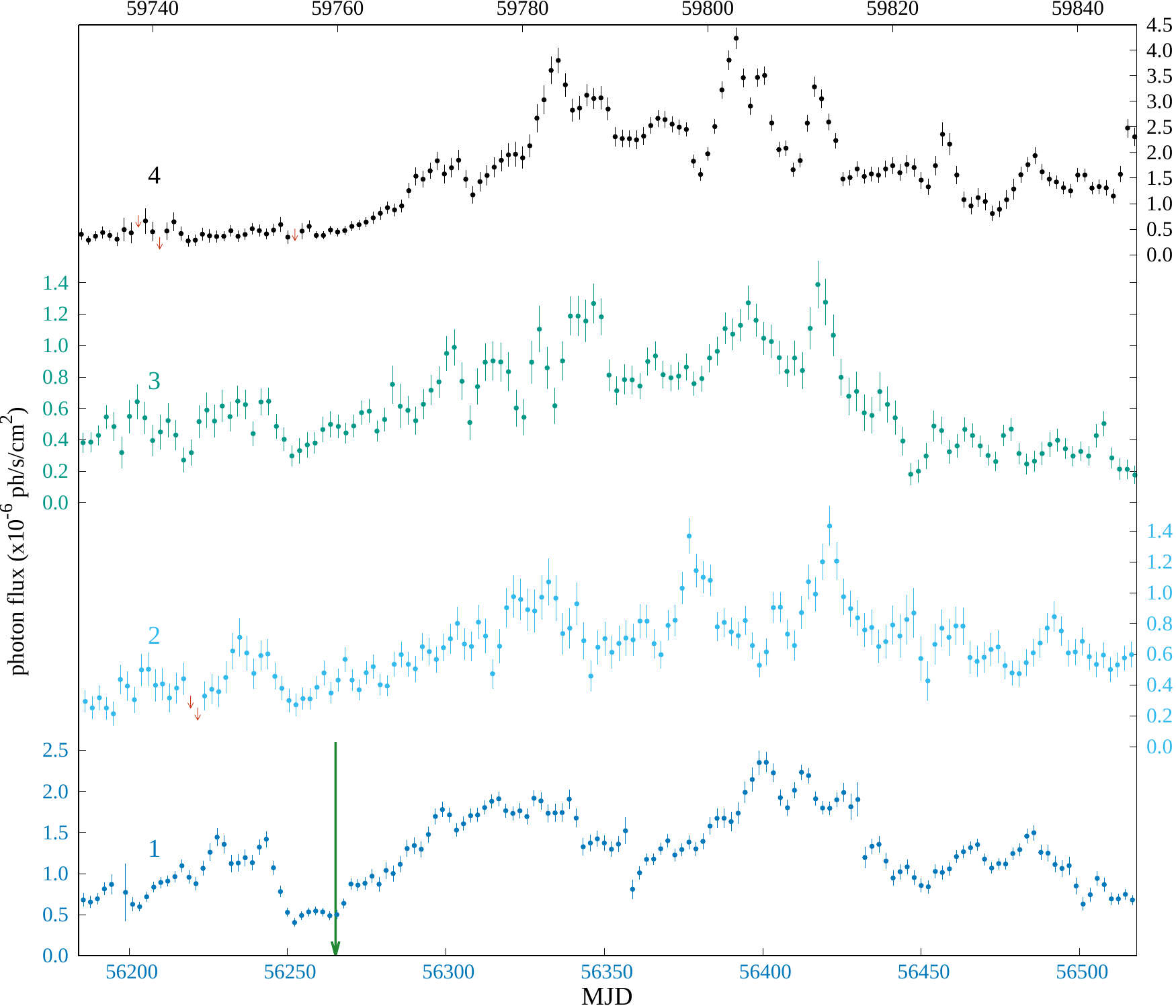}
    \caption{Repeated pattern of gamma-ray flares in 4FGL J1427.9$-$4206. The time scales of the curves 2, 3 and 4 are multiplied by factors 2.22, 2.08 and 2.94. The integration time for the curves is $t_{\rm int}=$ 4.4, 2.0, 2.35 and 1.54 d with $t_{\rm int}/2$ d between consecutive points. The green arrow indicates the arrival time of IC-35 "Big Bird" neutrino.}
    \label{fig:1427p1}
\end{figure*}

\begin{figure*}
	\includegraphics[width=0.8\textwidth]{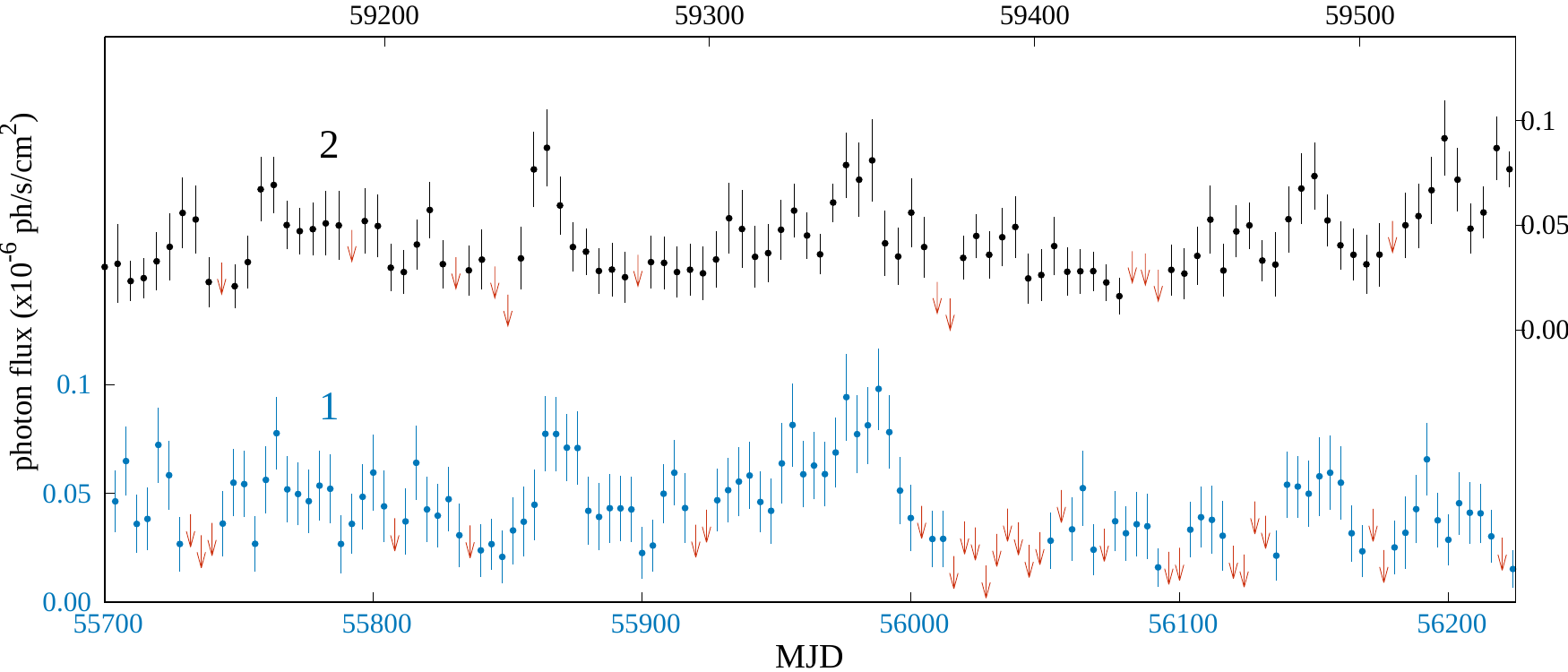}
    \caption{Repeated pattern of gamma-ray flares in 4FGL J1748.6+7005. The time scale of the curve 2 is stretched by a factor of 1.21. The integration time for both curves is $t_{\rm int}=8$ d with $t_{\rm int}/2$ days between consecutive points.}
    \label{fig:1748p1}
\end{figure*}

\begin{figure*}
	\includegraphics[width=0.8\textwidth]{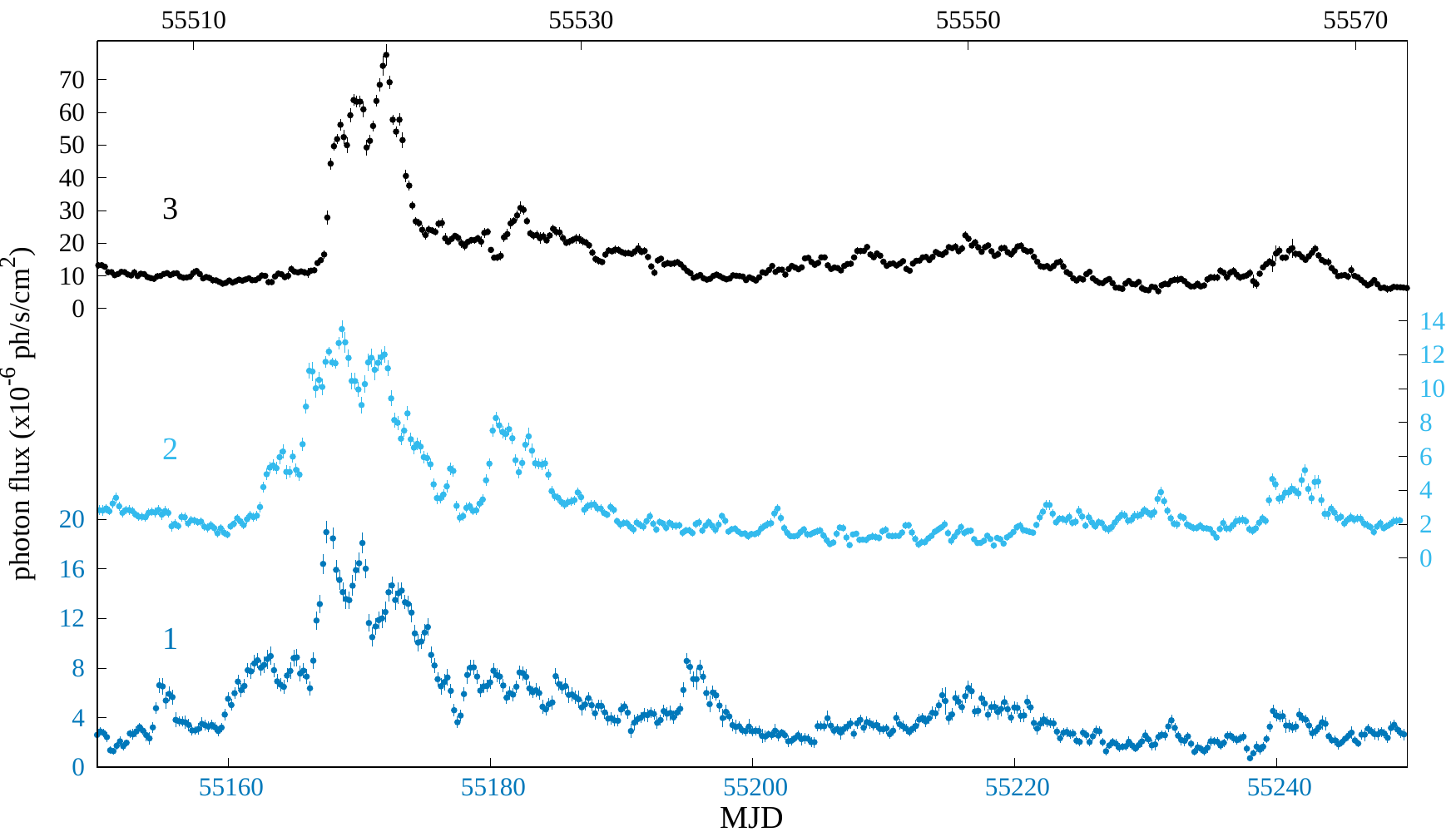}
    \caption{Repeated pattern of gamma-ray flares in 4FGL J2253.9+1609. The time scales of the curves 2 and 3 are squeezed and stretched by factors 0.44 and 1.48, respectively. The integration time for the curves is $t_{\rm int}=$ 0.5, 1.14 and 0.34 d with $t_{\rm int}/2$ d between consecutive points.}
    \label{fig:2253p1}
\end{figure*}

We found 22 repeated intervals of the light curves of 8 blazars from our sample, including the patterns in J1256.1$-$0547 (3C~279) and J2253.9+1609 (3C~454.3) reported earlier (see section~\ref{sec:intro}). The parameters of these repeated patterns, as well as the names and classes of the corresponding sources (hereafter "repeaters"), are listed in the top part of Table~\ref{tab:res}. The graphical representation of the found events is shown in Figures \ref{fig:0108p1} - \ref{fig:2253p1}, where the reference curve with id = 1 is always in its original time scale, while others are adjusted (stretched / compressed and shifted) according to the values listed in Table~\ref{tab:res}, which provide the minimum of TWED between the $i$-th and the first pattern repetition.

Two events, the third repetition of the pattern in J0457.0$-$2324 and the first repetition in J1427.9$-$4206, were missed by the automated search and were detected during the visual inspection of the light curves of these sources. For this reason, $\delta_i/\delta_1$ factors in Table~\ref{tab:res} for these sources are out of the range $[0.5,2]$ used for the automated pattern search. For J1256.1$-$0547, we adopted the pattern repetitions and their parameters from \cite{Blinov2021}. For J2253.9+1609, the pattern was identified by \cite{Jorstad2013}, where they were presented without any time scale modification. According to our analysis, adjusting the time scale by factors 0.44 and 1.478 in the second and the third events gives a better agreement between the curves.
\begin{table*}
 		\centering
 		\caption{Detected repeated patterns. (1) - Blazar id in 4FGL and an alternative name; (2) - synchrotron peak position class and type of the source; (3) - Compton dominance; (4) - identification number of the pattern; (5, 6) - start and end moments of the pattern; (7) - optimal time axis compression / stretching factor that provides the best match between the $i$-th and the 1-st repetition; (8) - way of identification of the repetition or reference where it is reported for the first time; (9) - probability of random similarity with the first repetition; (10) - joint significance of the pattern. }
 		\label{tab:res}
 		\begin{tabular}{lccccccccc}
 			\hline
 			Blazar & Class${}^a$  & CD${}^a$ & Repetition id. & MJD start & MJD end & $\delta_{i}/\delta_1$ & identification & $p-{\rm value}$ & S.L. \\
 			\hline
 			J0108.6+0134  & LSP fsrq & 6.2 & 1 & 59153 & 59410 & 1.00 & automated & - & $3.8\sigma$\\
 			  = 4C +01.02   &          &     & 2 & 59591 & 59737 & 1.759$\pm 0.002$ & automated & $1.4\times10^{-4}$ & \\
 			\hline
 			J0457.0$-$2324 & LSP FSRQ & 8.5 & 1 & 57547 & 57632 & 1.00 & automated & - & $4.8\sigma$\\
 		= PKS 0454$-$234   &           &     & 2 & 58645 & 58758 & 0.753$\pm 0.005$ & automated  & $1.3\times10^{-4}$ & \\
 			             &           &     & 3 & 59460 & 59499 & 2.18$\pm 0.02$ & visual & $9.7\times10^{-3}$ & \\
 			\hline
 			J1048.4+7143   & LSP FSRQ & 9.6 & 1 & 56550 & 56765 & 1.00 & automated & - & $3.4\sigma$\\
 			= S5 1044+71   &           &     & 2 & 57651 & 57866 & 1.000$\pm 0.005$ & automated & $6.4\times10^{-4}$ & \\
 			\hline
 			J1256.1$-$0547 & LSP FSRQ & 3.5 & 1 & 55002 & 55134 & 1.00 & \cite{Blinov2021} & - & $5.2\sigma$\\
 			= 3C 279       &           &     & 2 & 55672 & 55755 & 1.582$\pm 0.003$ & \ditto & $1.8\times10^{-3}$${}^c$ & \\
 			             &           &     & 3${}^b$ & 57280${}^b$ & 57516 & 0.560$\pm 0.003$ & \ditto & $1.7\times10^{-3}$${}^c$ & \\
 			               &           &     & 4${}^b$ & 56630${}^b$ & 56683 & 2.507$\pm 0.006$ & \ditto & $4.6\times10^{-2}$ & \\
 			\hline
 			J1345.5+4453   & LSP fsrq & 44.6 & 1 & 56587 & 56785 & 1.00 & automated & - & $3.1\sigma$\\
 		 = B3 1343+451	  &           &      & 2 & 56987 & 57185 & 1.000$\pm 0.002$ & automated & $1.5\times10^{-3}$ & \\
 			\hline
                J1427.9$-$4206 & LSP FSRQ & 3.5 & 1 & 56184 & 56518 & 1.00 & visual & - & $5.2\sigma$\\
 		= PKS 1424$-$41    &           &     & 2 & 57279 & 57429 & 2.216$\pm 0.001$ & automated & $5.1\times10^{-3}$ & \\
 			               &           &     & 3 & 57704 & 57865 & 2.08$\pm 0.01$ & automated & $1.2\times10^{-3}$ & \\
 			               &           &     & 4 & 59732 & 59846 & 2.92$\pm 0.01$ & automated & $3.0\times10^{-2}$ & \\
 			\hline
 			J1748.6+7005   & ISP bll  & 1.6 & 1 & 55700 & 56225 & 1.00 & automated & - & $2.2\sigma$\\
		    = S4 1749+70   &           &     & 2 & 59114 & 59547 & 1.21$\pm 0.01$ & automated & $2.7\times10^{-2}$ & \\
		    \hline
		    J2253.9+1609   & LSP FSRQ & 6.4 & 1 & 55150 & 55250 & 1.00 & \cite{Jorstad2013} & - & $5.0\sigma$\\
		     = 3C 454.3       &           &     & 2 & 55252 & 55479 & 0.4400$\pm 0.0002$ & \ditto & $6.2\times10^{-4}$ & \\
		                   &           &     & 3 & 55505 & 55573 & 1.478$\pm 0.003$ & \ditto & $8.9\times10^{-4}$ & \\
                \hline
                \multicolumn{10}{c}{Additional visual search among neutrino blazar candidates}\\
                \hline
     		J0505.3+0459  &  LSP FSRQ & 3.1  & 1 & 56820 & 56907 & 1.00 & visual & -  & $3.3\sigma$\\
 		 = PKS 0502+049  &           &      & 2 & 57020 & 57090 & 1.237$\pm 0.006$ & visual & $2.2\times10^{-2}$ & \\
 			               &           &      & 3 & 57538 & 57635 & 0.90$\pm 0.02$ & visual & $4.2\times10^{-2}$ & \\
                \hline
                J0509.4+0542  & ISP BLL  & 0.7   & 1 & 55510 & 55670 & 1.00 & visual & - & $3.0\sigma$\\
 		   = TXS 0506+056  &          &       & 2 & 57871 & 58049 & 0.899$\pm 0.006$ & visual & $2.0\times10^{-3}$ & \\
                \hline
		    \multicolumn{10}{l}{${}^a$ as reported in the 4LAC DR3 \citep{Ajello2022}: FSRQ and fsrq are identified and associated Flat Spectrum Radio}\\
		    \multicolumn{10}{l}{Quasars; BLL and bll are identified and associated BL Lacertae objects. LSP and ISP stand for low- and intermediate-synchrotron-peaked sources.}\\
                 \multicolumn{10}{l}{${}^b$ not in the chronological order to be consistent with the numeration in \protect\cite{Blinov2021}. ${}^c$ adopted from \protect\cite{Blinov2021}.}\\
 		\end{tabular}
\end{table*}

In order to assess significance of the found patterns, we employed the procedure used in \cite{Blinov2021}. We estimated the probability that the i-th repetition of a pattern under the found transformation only accidentally resembles the first repetition using the Euclidean distance \citep[ED,][]{Lin2012} as a measure of similarity. For this purpose, we performed a Monte-Carlo simulation that selected a random piece of the entire light curve and a random time scale transformation factor. After transforming the selected interval timescale using this factor and shifting it to the Julian Date range of the first pattern repetition, we calculated the ED between the two light curve intervals. The photon fluxes of the two light curve intervals used in this procedure were normalised to be in the range [0, 1]. The timescale transformation factor was allowed to vary in the range [0.5, 2] as in the patterns search procedure for all cases, except the events where $\delta_{i}/\delta_1$ is outside this range (see Table~\ref{tab:res}). For the latter cases, we extended the allowed range to [$\delta_1/\delta_{i}$ - 0.1, $\delta_{i}/\delta_1$ + 0.1]. By performing $N=10^6$ simulations, we determined the number of trials M where the ED value was smaller than that of the detected pattern, indicating that a random light curve interval is closer (more similar) to the first pattern repetition. We then computed $p-{\rm value} = \frac{M + 1}{N + 1}$ \citep{Davison1997}, which estimates the randomness of similarity between each pattern's first and i-th repetitions. These values are listed in Table~\ref{tab:res}. For most of the repetitions, the $p-{\rm values}$ are below $2.7\times10^{-3}$, implying $>3\sigma$ significance. In cases with more than two repetitions, the joint significance of the pattern is provided by multiplying the $p-{\rm values}$ for all repetitions. For instance, in the case of J1427.9$-$4206, the overall significance is $p-{\rm value} = 1.8\times 10^{-7}$ ($5.2\sigma$). We provide the joint significance level for each pattern in the last column of Table~\ref{tab:res}. These values reveal that the majority of the identified patterns are unlikely to be random.

It is worth noting that the procedure used in this study for identifying repeated patterns is only sensitive to complex light curve profiles with multiple flares on the time scale of tens to hundreds of days. There is a possibility and physical justification \citep{Shukla2020} for more trivial repeated single flares with much shorter duration, but we would not be able to detect such events. Additionally, as seen in Figures \ref{fig:0108p1} - \ref{fig:2253p1}, the found repeated patterns are far from identical, with individual flares slightly changing their relative amplitude from event to event. In some cases, a shift in time of a particular flare in the pattern with respect to others is also present. Our sensitivity to such cases is limited, and it is likely that we missed a number of repeated patterns in other sources. We estimate that the loss rate is not less than $2/15 = 13$\%, as two out of 15 new patterns found in this work were missed by the automated algorithm. Therefore, currently we can only provide a lower limit on the occurrence of patterns based on our flux-limited sample. Our search shows that at least 8 per cent of gamma-ray sources exhibit repeated patterns of flares in their light curves.

\section{Neutrino association} \label{sec:neutr}

\cite{Tavecchio2014} explored the possibility that structured jets could provide a mechanism for efficient neutrino production in AGN. They showed that a boost of 2 - 3 orders of magnitude in neutrino luminosity, similar to the increase observed in inverse Compton gamma-ray emission \citep{Ghisellini2005}, could occur for the spine's neutrino output when a sheath around it is introduced. This boost is a result of high energy protons undergoing photo-meson reactions with the amplified soft target photon field of the sheath. If the patterns of gamma-ray flares found in section \ref{sec:res} are produced due to the presence of a sheath in the jet (which we discuss in section \ref{sec:mech}), one would expect to observe a correlation between the patterns and the neutrino flux if the model proposed by \cite{Tavecchio2014} is correct. 

We investigated this hypothesis using neutrinos detected in different experiments and possible neutrino associations with individual blazars reported in the literature \citep[e.g.,][]{Boettcher2022}. We found that three out of the eight repeaters could be associated with positionally consistent neutrinos that arrived during the time intervals of repeated patterns. We have marked these three neutrino events in Figures \ref{fig:0108p1}, \ref{fig:1256p1} and \ref{fig:1427p1}, as well as in the entire available light curves in Figures \ref{fig:0108full}, \ref{fig:1256full} and \ref{fig:1427full}, and discuss them below.

Neutrino GVD210710CA was detected by Baikal-GVD and arrived on MJD = 59405.6 from a direction that includes J0108.6+0134 within the 90\% containment area \citep{Allakhverdyan2022}. The moment of arrival is close to the end of the first repetition of the pattern we detected in this source. Although this neutrino has a relatively low energy 24.5 TeV, it belongs to the category of under-horizon events that have >70\% probability of being astrophysical \citep{Allakhverdyan2022}.

The second possible association is with a neutrino detected on MJD = 57291.9 and reported as an Extremely High Energy alert-like (EHEA) event by the IceCube collaboration. This neutrino, called IC150926A in the ICECAT-1 catalogue, had an energy of $216$ TeV \citep{Abbasi2023}. According to the analysis by \cite{Plavin2020}, this neutrino is likely to be associated with J1256.1$-$0547 (3C 279). Indeed, the blazar is located within the 90\% containment area of the neutrino, while its arrival time is within the third repetition of the pattern found in \cite{Blinov2021} and listed in Table~\ref{tab:res}.

The third neutrino coincident in time with a pattern and in arrival direction with the corresponding repeater is the third PeV neutrino detected by the IceCube and dubbed as IC-35 ("Big Bird"). A possible association of this particle with J1427.9$-$4206 (PKS B1424$-$418) is discussed by \cite{Kadler2016}. It arrived at MJD = 56265.1 during a major gamma-ray outburst, which is found to be the first repetition of the pattern found in the light curve of this blazar.

\begin{figure*}
	\includegraphics[width=0.8\textwidth]{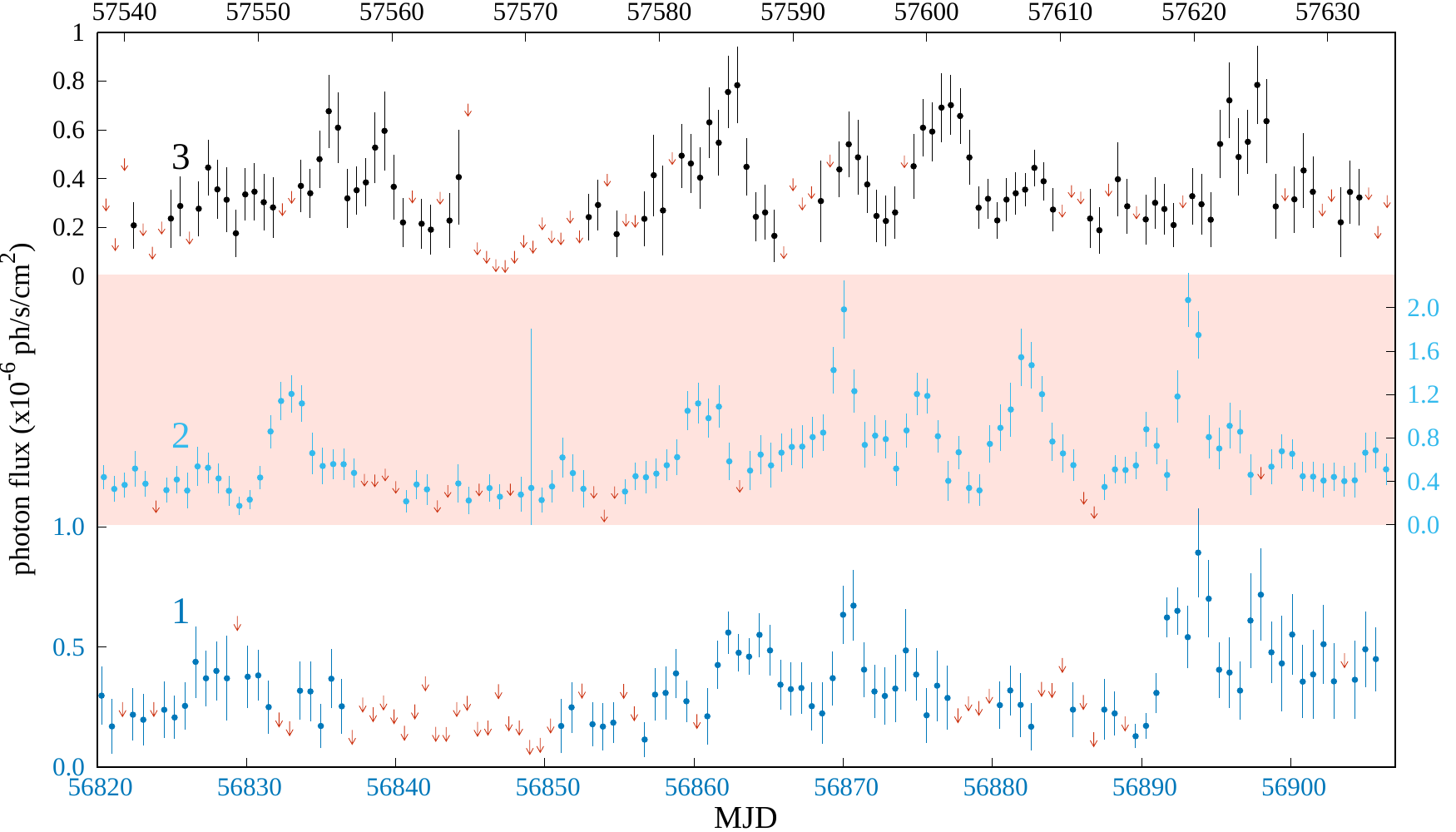}
    \caption{Repeated pattern of gamma-ray flares in 4FGL J0505.3+0459. The time scales of the curves 2 and 3 are stretched and squeezed by factors 1.24 and 0.9, respectively. The integration time for the curves is $t_{\rm int}=$ 1.4, 1.13 and 1.39 d with $t_{\rm int}/2$ d between consecutive points. The pink region marks the $\sim 3.7\sigma$ significance neutrino excess interval from MJD=56937.8 to MJD=57096.2 reported in \protect\cite{Aartsen2018b}.}
    \label{fig:0505p1}
\end{figure*}

\begin{figure*}
	\includegraphics[width=0.8\textwidth]{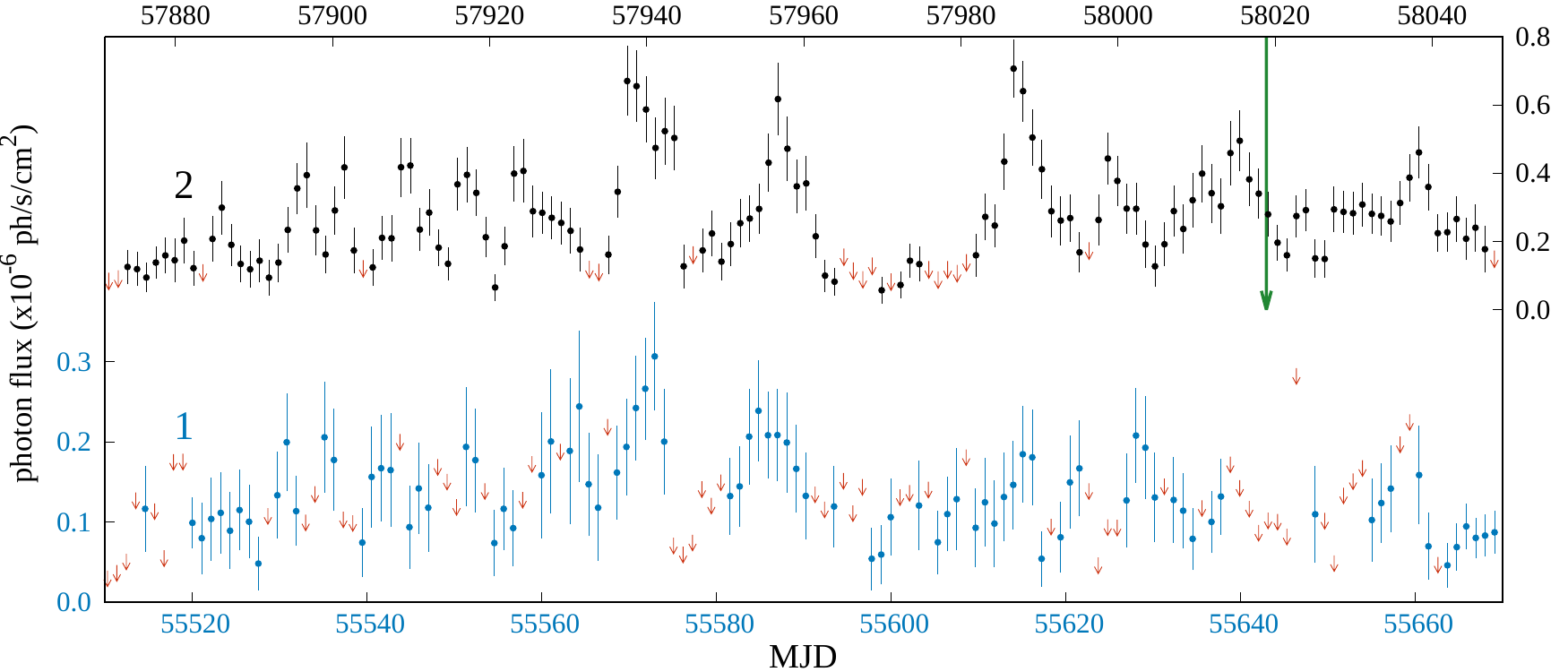}
    \caption{Repeated pattern of gamma-ray flares in 4FGL J0509.4+0542. The time scale of the curve 2 is squeezed by a factor of 0.9. The integration time is 2.2 and 2.4 d for curves 1 and 2, respectively. The time difference between consecutive points is 1/2 of the bin duration. The green arrow indicates the arrival time of the IC170922A neutrino.}
    \label{fig:0509p1}
\end{figure*}

Since there are only a few blazars that have been tentatively associated with neutrinos, where this association is studied at least in some detail in the literature, we conducted a visual search for patterns in the light curves of these sources. Out of the six blazars (excluding J1427.9-4206) discussed as possible neutrino sources \citep{Boettcher2022}, only four J0505.3+0459 (PKS 0502+049), J0509.4+0542 (TXS 0506+056), J0738.1+1742 (PKS 0735+17), and J1504.4+1029 (PKS 1502+106) have high enough gamma-ray photon fluxes for the repeated pattern search. All these four blazars are in the top 100 brightest sources of 4FGL, i.e. our sample, for which we analysed the gamma-ray light curves with 3 and 7 days binning. However, only J1504.4+1029 is in the top 30 list for which we processed and analysed also the one day binned light curves (see section~\ref{subsec:gamma}). For this reason, repeated patterns could have been overlooked by our algorithm in the three blazars. Due to the low average photon flux from these sources, the light curves with one day binning had an insufficient number of flux detections. Therefore, for J0505.3+0459, J0509.4+0542 and J0738.1+1742 we processed the {\em Fermi}-LAT data in the same way as for the top 30 sample, but with two days integration time. By visually inspecting these light curves we identified two repeated patterns of flares in J0505.3+0459 and J0509.4+0542 that are listed at the bottom of Table~\ref{tab:res} and are shown in Figs.~\ref{fig:0505p1} and \ref{fig:0509p1}. The complete light curves of these sources are shown in Figs.~\ref{fig:0505full} and \ref{fig:0509full}. The pattern in J0509.4+0542 does not follow the dependencies of the flux on the time scale (or $\delta$) transformation we considered in section~\ref{sec:meth}. The second repetition of the pattern lasted 1.1 times longer compared to the first, which means that $\delta$ was lower for this event. However, contrary to our expectation, fluxes near peaks of the two major flares of this repetition are $>2$ times higher compared to the first repetition. Same inconsistency between the data and the assumption used in our search procedure is found for the first and third repetitions of the pattern in J0505.3+0459. This presumably means that our model of flux behaviour during repeated patterns is oversimplified. We discuss this possibility further in section~\ref{sec:disc}. On the other hand, it means that we probably miss many other repeated patterns that do not follow our assumptions using the automated search as described before.

J0509.4+0542 is the first blazar where a high-confidence association of a multi-band flare with a neutrino detection was discovered \citep{Aartsen2018a}. It is associated with the $E = 264$ TeV neutrino IC170922A, which arrived on MJD=58018.9.
This is currently the most extensively studied connection between a neutrino and an AGN \citep[e.g.,][and references therein]{Halzen2022}, and the most reliable to date. After the first association with IC170922A, the IceCube collaboration conducted a retrospective analysis of their data. They found an excess of high-energy neutrino events, with respect to the atmospheric background, at the position consistent with the blazar between September 2014 and March 2015 \citep{Aartsen2018b}. However, during this time interval, the blazar was in a very low state in all energy bands, including gamma-rays (see Fig.~\ref{fig:0509full}). This led to an extended discussion in the literature, suggesting that the high energy neutrino flux is not necessarily correlated with gamma-ray emission at GeV energies and higher. The reason for this is that the opacity $\tau_{p\gamma}$ of $p\gamma$ interactions is much smaller than the opacity $\tau_{\gamma\gamma}$ for $\gamma\gamma$ interactions. In a zone with efficient neutrino production $\tau_{p\gamma} \gg 1$, and as a consequence, the gamma-rays are unable to escape such a zone \citep[e.g.,][]{Reimer2019}. However, an alternative interpretation has been proposed for this set of neutrino events previously associated with J0509.4+0542. \cite{Sumida2022} studied the kinematics of individual radio knots in this source and the nearby blazar J0505.3+0459 using MOJAVE data. They found that the excess of neutrinos in 2014 - 2015 from the direction consistent with both blazars is coincident in time with the ejection of a new radio knot in J0505.3+0459, which was accompanied by the two strongest gamma-ray flares in this source (see Fig.~\ref{fig:0505full}). Therefore, \cite{Sumida2022} conclude that there is a possibility that at least some neutrinos of the excess in 2014 - 2015 arrived from J0505.3+0459 instead of J0509.4+0542. A similar interpretation has also been proposed in other works \citep{He2018,Liang2018,Banik2020}. The repeated patterns found in these two blazars support this scenario. IC170922A arrived within the time interval of the repetition 2 of the pattern in J0509.4+0542 (see Figs. \ref{fig:0509p1} and \ref{fig:0509full}). On the other hand, the excess of neutrinos in 2014 - 2015 is consistent with the first two repetitions of the pattern in J0505.3+0459 (see Fig.~\ref{fig:0505full}). We note that in the latter figure, we highlight the box-shaped interval of the highest significance of the excess reported in \cite{Aartsen2018b}, while the $3\sigma$ significance interval spans earlier in time (until the middle of 2014) and covers the first repetition of our pattern entirely.

It appears unlikely that the occurrence of the four individual neutrino events, together with the excess of neutrinos, in coincidence with the repeated patterns in both arrival time and direction can be attributed to mere chance. However, it is hard to estimate the statistical significance of the link between these events given the very different properties of the cascade- and the track-like events, as well as the dissimilarities between the IceCube and Baikal-GVD detectors. Therefore, to assess the significance of the association, we only use the uniformly selected sample of 54 track-like IceCube neutrinos described in section \ref{subsec:neutr}. We follow the procedure used by \cite{Plavin2020} and \cite{Hovatta2021}. For each neutrino in the sample, we translate the asymmetric coordinate-wise uncertainties of the arrival direction provided in ICECAT-1 into two-dimensional coverage regions. In this operation, the positive and negative uncertainties $\Delta^{\rm stat}_i$ of the right ascension (RA) and the declination (DEC) are multiplied by $\frac{\sqrt{-{\rm log}(1-0.9)}}{{\rm erf}^{-1}(0.9)} \approx 1.3$, as described by \cite{Plavin2020}. Aside from the statistical errors, an extra systematic positional uncertainty $\Delta \Psi$ exists, primarily governed by the uncertainties in the optical properties of the ice \citep{IceCube2013}. The values of $\Delta \Psi$ are not available for all neutrino events. Therefore, following \cite{Plavin2020} and \cite{Hovatta2021}, we apply a procedure often used in particle physics \citep[see e.g.,][]{Abbasi2012}, where the signal is maximized by tuning the unknown parameter. This procedure requires a multiple testing correction that adjusts the derived p-value. We consider a single value of $\Delta \Psi$ for all events and perform a scan across the range of $\Delta \Psi$ from $0.1\dg$ to $1.0\dg$ with an increment of $0.01\dg$. Each of the statistical uncertainties $\Delta^{\rm stat}_i$ is propagated in quadrature with the systematic uncertainty $\Delta \Psi$, resulting in a total uncertainty for each neutrino of $\sqrt{(\Delta^{\rm stat}_i)^2 + (\Delta \Psi)^2}$. The four derived uncertainty values are used pairwise to construct four quarters of ellipses that delineate the region of uncertainty  for the arrival direction of each neutrino. We define our statistic $\nu$ as the number of blazars falling into the uncertainty regions of the neutrinos, and at the same time, having an ongoing repeated pattern during the corresponding neutrino arrival moment. For each $\Delta \Psi$ we first find the value of $\nu_{\rm obs}$ using the observed coordinates of neutrino events. Then we perform a Monte Carlo (MC) simulation assigning a random RA while keeping the real DEC and arrival time for each neutrino in the sample. The identical method was used in previous studies \citep[e.g.,][]{Plavin2020,Hovatta2021}. This way of shuffling is suitable for the IceCube events because its sensitivity to high accuracy depends only on the zenith angle \citep{Aartsen2017b}, while the telescope is located at the south pole. Therefore, by shuffling only RA values, we effectively move the detected neutrino events along its isosensitivity curves, avoiding the need to introduce any correction for the non-uniform response of the detector across the sky. We also randomized the intervals of repeated patterns, placing them at random positions in time along the analyzed light curves, while maintaining the observed duration. Performing $N=10^5$ MC trials, we compute the same statistic $\nu_i, 1 \le i \le N$ for the randomly shifted neutrinos. We then determine the number of trials M where $\nu_i \ge \nu_{\rm obs}$. After this we calculate the probability of $\nu_{\rm obs}$ being random as \citep{Davison1997}:
\begin{equation} \label{eq:1}
{\rm p-value} = \frac{M+1}{N+1}.
\end{equation}
This p-value depends on the value of the $\Delta \Psi$, which is a free parameter. Varying $\Delta \Psi$ across the range of interest and minimizing the p-value, we find the pre-trial p-value. In our case the pre-trial ${\rm p-value} = 1.2 \times 10^{-4}$ is reached for $\Delta \Psi = 0.99\dg$ and $\nu_{\rm obs} = 2$, which is provided by IC170922A and IC150926A associated with J0509.4+0542 and J1256.1$-$0547, respectively. To correct for the multiple trials, we conduct an additional layer of MC simulations using the sets of generated neutrinos and the corresponding $\nu_i$ values. Each random sample is treated as an actual observation. It is compared with $N=99999$ other random samples, and the corresponding p-value is calculated for each $\Delta \Psi$. As with the real data, we find the pre-trial p-value for each random sample by minimizing across the range of $\Delta \Psi$. Subsequently, the number of simulations M is calculated, wherein the pre-trial p-value is either smaller or equal to the one obtained from the observed neutrino sample. Using equation~\ref{eq:1} we find the post-trial ${\rm p-value} = 4.2 \times 10^{-3}$ ($2.8\sigma$) which is unaffected by the multiple comparisons issue. This value provides a significance estimate for the association of neutrinos with repeated patterns.

\section{Properties of repeaters} \label{sec:prop}

It is important for the interpretation of the found patterns to understand whether the blazars producing them are somehow different from "non-repeaters", which are blazars that do not exhibit patterns.

We verified whether all ten identified repeaters stand out from non-repeaters in different properties listed in the Fourth LAT AGN Catalog \citep[4LAC DR3,][]{Ajello2022}. Using the two-sample Kolmogorov-Smirnov (KS) test, we examined the null hypothesis that the two samples were drawn from the same distribution, considering various properties of blazars. We found that the redshift distributions for the two classes are consistent with the null hypothesis ${\rm p-value}=0.3$. The distributions of the photon index of the Power-Law fit to the gamma-ray spectral energy distribution (SED) of sources are consistent for the two samples ${\rm p-value}=0.4$. Repeaters have a tendency to have higher luminosity compared to non-repeaters, which, however, is found to be insignificant ${\rm p-value}=0.2$. The Compton dominance (CD), defined as the ratio between the peak $\nu F_{\nu}$ for the high- and low-frequency SED components, tends to be higher for repeaters, but insignificantly $p-{\rm value}=0.018$ ($2.4\sigma$). Another parameter that has a tendency to be higher for repeaters, but not significantly according to the KS test $p-{\rm value}=0.009$ ($2.6\sigma$), is the variability index provided in the 4LAC.

\begin{figure}
\centering
  \includegraphics[width=0.7\columnwidth]{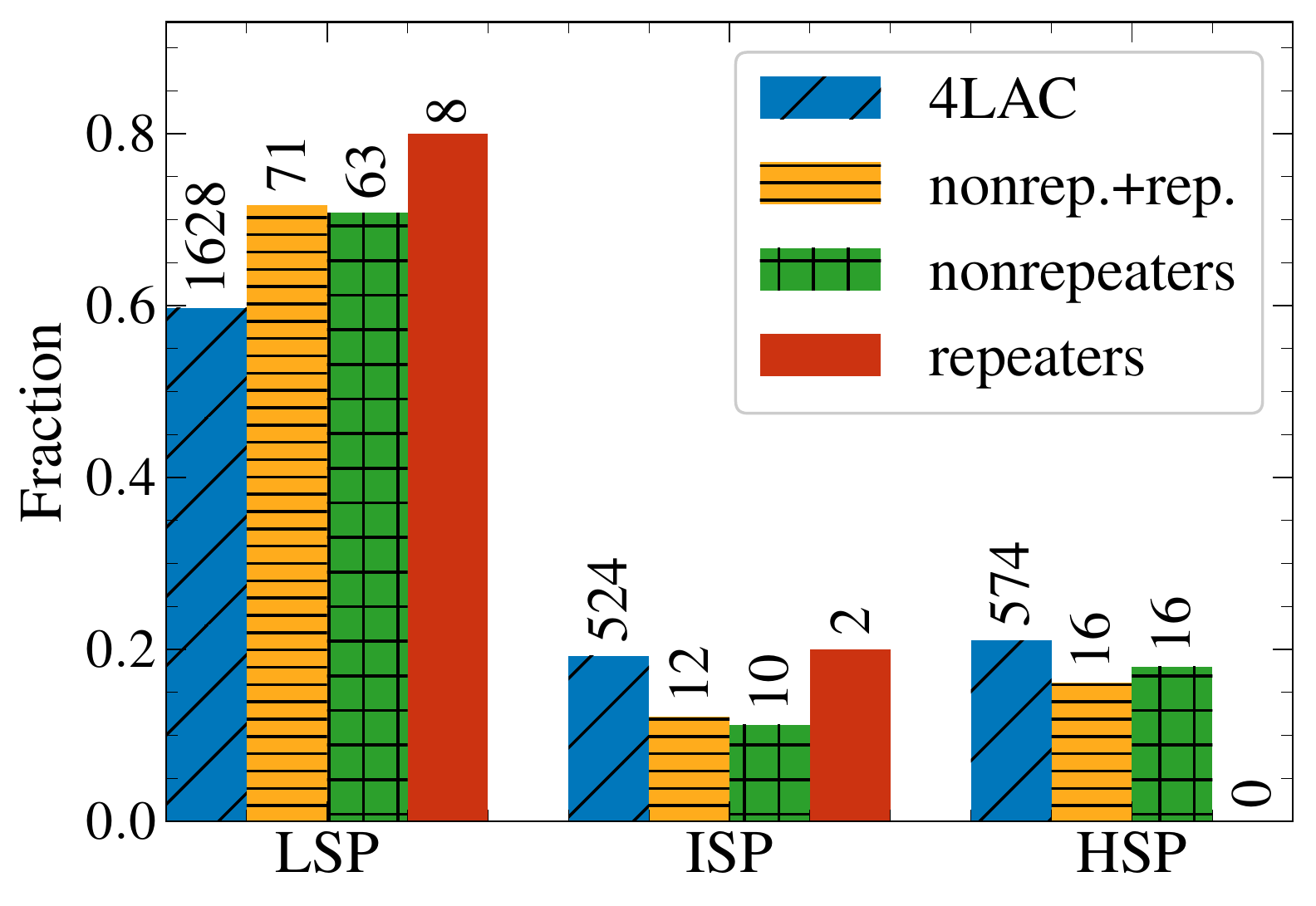}
    \caption{Distribution of sources in different samples by the synchrotron peak class.}
    \label{fig:spcl}
\end{figure}
We also studied how sources are distributed among the synchrotron peak position classes. In Figure~\ref{fig:spcl}, we present a histogram in which all sources are separated into low- (LSP), intermediate- (ISP) or high-synchrotron peak (HSP) blazars. For all four considered samples: the entire 4LAC; our flux limited sample of 100 brightest blazars; repeaters, and non-repeaters, LSP sources strongly dominate among others. Among the sample of repeaters, this domination is even more prominent -- 8 out of 10 repeaters are LSP blazars. Moreover, the same 8 repeaters belong to Flat Spectrum Radio Quasars (FSRQ) class. Only two sources, J0509.4+0542 and J1748.6+7005, belong to the ISP and to BL Lacertae (BLLacs) objects. Although, according to \cite{Padovani2019}, J0509.4+0542 is a masquerading BL Lac, i.e. intrinsically a flat-spectrum radio quasar with hidden broad lines. In either case, this distribution of repeaters among blazar classes can be a chance coincidence. We found that such sample of repeaters as the one we discovered can be randomly selected from our initial sample of 100 sources with a probability $\sim 5$ per cent for the distribution by the synchrotron peak frequency and $\sim 6$ per cent (or $\sim 2$ per cent if J0509.4+0542 indeed belongs to FSRQ) for the distribution by FSRQ/BLLacs class.

We also compared the distribution of the Doppler factor of sources, taken from \cite{Liodakis2018}, for the two samples. The KS test showed that the null hypothesis of  identical parent distribution can not be rejected ${\rm p-value}=0.8$. Additionally, we collected median apparent velocities of radio knots as reported by the MOJAVE collaboration \citep{Lister2019} and translated them into the intrinsic
jet speed $\beta = v/c$, using the jet viewing angle derived from variability \citep{Liodakis2018}. The distributions of $\beta$ for repeaters and non-repeaters do not differ significantly according to the KS test ${\rm p-value}=0.9$, although the sample sizes are significantly reduced due to missing data. These tests were motivated by the theoretical possibility of a so-called Compton rocket effect taking place in structured jets \citep{Sikora1996,Ghisellini2005}. If the repeated patterns are caused by the presence of sheaths in the jets, which provide seed photons for the EC mechanism occurring in emission features travelling in the spine, then these emission features are expected to decelerate due to the high anisotropy of the EC emission. The deceleration can potentially be observed in the kinematics properties of repeaters compared to non-repeaters.

It has been suggested that the SED of a jet sheath peaks in the infrared band \citep{Marscher2010}. In order to verify whether repeaters have an excess or unusual colours in far or near infrared with respect to non-repeaters we collected WISE \citep{Wright2010} and 2MASS \citep{Skrutskie2006} data for sources in our sample.
The distribution of all sources on the W1-W2 vs. W2-W3 plane is well consistent with the WISE gamma-ray strip \citep{Massaro2011}. The sample of repeaters, being dominated by FSRQ, is redder than the entire sample of non-repeaters. However, when repeaters are compared only to non-repeaters of the FSRQ class, there is no significant difference in colours. The KS test p-values for W1-W2 and W2-W3 distributions of the two samples are 0.5 and 0.6. All repeaters except J1256.1-0547 are located in the region of the colour-colour space where most of the FSRQ lie \citep[see Fig.~4 of][]{Massaro2012}. Near infrared colours J-H and H-K obtained from 2MASS data have very similar distributions for repeaters and non-repeaters. Therefore, we conclude that blazars that exhibit repeated patterns do not show peculiar signatures in the infrared part of the SED. However, this does not reject the possibility that the patterns are caused by the presence of a sheath in the jet. \cite{MacDonald2015} estimated that a sheath can provide enough seed photons for strong EC gamma-ray flares even when its bolometric luminosity in the observer's frame is an order of magnitude smaller than the total jet luminosity.

\section{Mechanism behind the patterns} \label{sec:mech}

As we discussed in the introduction, there are two models proposed for the explanation of irregularly spaced repeated patterns. In both of these models, the patterns are caused by emission features propagating in the jet, although the emission mechanism in flares is different. In the first scenario, individual flares of a pattern are caused by the interaction of the moving component with stationary recollimation shocks in the jet. In this case, the emission during flares has a synchrotron self-Compton origin \citep{Jorstad2013, Hervet2019}. In the second scenario, the moving emission feature propagates in the fast spine of the jet through a system of quasi-stationary ring-like overdensities of plasma in the slower sheath around the spine. Since the emission of rings is relativistically boosted in the reference frame of the moving feature and vice versa, this causes a burst of efficient inverse Compton up-scattering of the external seed photons to gamma-ray energies \citep{Ghisellini2005}. In this section we discuss observational evidence that can discriminate between the two scenarios.

The suggested models imply the propagation of emission features in the jet, which can potentially be detected as moving knots with Very Long Baseline Interferometry (VLBI) experiments. Therefore, we conducted a search in the literature for radio-knots kinematics data to verify their potential association with the identified repeated patterns. Among the 10 repeaters, only four sources (J0505.3+0459, J0509.4+0542, J1256.1$-$0547, and J2253.9+1609) have been extensively studied using VLBI. We show the moments of radio-knots ejections in these sources found in the literature in Figures \ref{fig:1256full}, \ref{fig:2253full}, \ref{fig:0505full}, and \ref{fig:0509full} and discuss these data further.

The radio knots kinematics data for J1256.1$-$0547, as presented by \cite{Jorstad2017}, were analysed in conjunction with repeated patterns in \cite{Blinov2021}. The analysis demonstrated that the pattern repetitions 1, 3 and 4 in this source are likely associated with radio knots ejections C29 or C29A, C34, and C36, as shown in Fig.~\ref{fig:1256full}. For the repetition 2, a knot was most likely overlooked due to presence of two very bright components in the vicinity of the radio core during its ejection \citep{Blinov2021}.

For J2253.9+1609, two radio knots, K09 and K10 in Fig.~\ref{fig:2253full}, were presented in the same paper where the pattern in this source was discovered \citep{Jorstad2013}. Following their work, we associate K09 with the pattern repetition 1 and K10 with repetition 3. The ratio of Doppler factors, $\delta_{K10}/\delta_{K09} = 1.9 \pm 0.3$, is consistent with $\delta_3/\delta_1 = 1.478\pm0.003$ within the $2\sigma$ confidence interval. An extended data set of the same 43 GHz VLBI monitoring program as in \cite{Jorstad2013} was analysed with a different approach by \cite{Weaver2022}. They report multiple knots ejections in this blazar, which we show in Fig.~\ref{fig:2253full}. Two of these knots, B10 and B11, emerged from the radio core at movements close to the time intervals of our patterns. The first of these knots, B10, is consistent in the ejection time and the Doppler factor with K09. However, for B11 it is unclear from the public data whether it is a revised counterpart of K10 or a completely different moving feature that should be associated with the second repetition of the pattern. The ejection times and the Doppler factors of B11 and K10 are significantly different. In either case, $\delta_{B11}/\delta_{B10} = 0.9 \pm 0.1$ is inconsistent with our $\delta_2/\delta_1 = 0.44$ and $\delta_3/\delta_1 = 1.478$, while it is in a good agreement with a constant $\delta$ initially suggested for the patterns in this source in \cite{Jorstad2013}. A detailed joint analysis of VLBI kinematics data and gamma-ray light curves for this sources could possibly better clarify the interpretation of these events.

The kinematics of radio knots in J0505.3+0459 and J0509.4+0542 has been studied by \cite{Sumida2022}. We show their ejection moments with the red arrows in Figs.~\ref{fig:0505full} and \ref{fig:0509full}. The patterns 1 and 2 of J0505.3+0459 are presumably associated with components C3 and C4. \cite{Sumida2022} do not provide Doppler factor estimates for detected radio knots. However, they list apparent velocities $\beta_{\rm app}$ of the components. Since $\beta_{\rm app}$ is approximately proportional to $\delta$, we use these values for comparison with our $\delta_i/\delta_1$. Specifically, $\beta_{\rm app, C4} / \beta_{\rm app, C3} = 2.2\pm1.4$ is consistent with $\delta_2/\delta_1 = 1.24$.  The first pattern repetition in J0509.4+0542 is presumably related to the knot C4, while the second pattern could be related to C9 or C10, both of which were ejected during the pattern time interval when the uncertainties are taken into account. The values $\beta_{\rm app, C10} / \beta_{\rm app, C4} = 0.6\pm0.9$ and $\beta_{\rm app, C9} / \beta_{\rm app, C4} = 1.3\pm0.3$ are both consistent with $\delta_2/\delta_1 = 0.899\pm0.006$ within the $2\sigma$ confidence interval, while C10 provides a better agreement.

For J1427.9-4206, we could not find any VLBI data reported in the literature. The remaining five sources (J0108.6+0134, J0457.0-2324, J1048.4+7143, J1345.5+4453, and J1748.6+7005) are part of the MOJAVE sample\footnote{\url{https://www.cv.nrao.edu/MOJAVE/allsources.html}}. However, none of them were observed with a sufficient cadence that would ensure the possibility of detecting new radio components adjacent to the time intervals of the repeated patterns. Therefore, the absence of new radio knots ejections in these five sources in MOJAVE data \citep{Lister2019} cannot be considered as evidence against the suggested models. Moreover, even in cases with sufficient sampling, VLBI monitoring data do not provide a complete sample of ejected radio knots \citep[e.g., see the discussion on the missing knot for the second repetition in J1256.1$-$0547 by ][]{Blinov2021}. Finally, the moving emission feature responsible for generating repeated patterns in the inner part of the jet may dissipate before reaching the radio core, rendering it undetectable through VLBI.

Many of the found repeated patterns have non-trivial profiles. For instance, the patterns in J0108.6+0134 and J1427.9-4206 appear to be a superposition of smooth, long-term and strong, sharp flares. The patterns in J1048.4+7143 and J1345.5+4453 are composed of multiple fast flares on top of a $\sim100$ d long period of elevated flux. It is hard to explain how such complex profiles, and especially, long duration, smooth underlying flares, can be produced by shock interactions with recollimation zones. This scenario is expected to provide isolated, short flares with a characteristic profile that has a fast rise and longer decay time \citep[e.g.,][]{Saito2015}. On the other hand, in the spine-sheath scenario, the patterns can have any profile determined by the longitudinal distribution of plasma in the sheath.

The repeated pattern in J1256.1$-$0547 is found to be associated with periods of long and smooth optical polarization plane rotations \citep{Blinov2021}. In particular, the second repetition of the pattern in this blazar is nearly perfectly coincident with a $\sim350\dg$ rotation of polarization \citep[discussed earlier in][]{Kiehlmann2016,Larionov2020}. If individual flares of this pattern are produced by an emission feature passing through separate recollimation zones, it is hard to explain the coherent smooth optical polarization behaviour along the entire time interval. One would expect that during each such recollimation zone crossing, the magnetic field is highly compressed and aligned with the shock front \citep{Marscher2010}. This should cause strong polarized flux flares along with a fixed polarization plane orientation, while the observed behaviour is inconsistent with this picture. We found that the beginning of the first pattern in J1048.4+7143 was also accompanied by a polarization plane rotation reported in \cite{Blinov2015}. However, the end of this rotation is not defined. Due to the lack of data, it is unclear whether it continued during the major flares of this event.

The SSC and EC mechanisms have distinct signatures when a very broad band SED is considered. For instance, if the gamma-ray emission is produced by the SSC mechanisms, Compton Dominance (CD), the ratio of the inverse Compton and the synchrotron flux, cannot greatly exceed unity \citep{Sikora2009}. Otherwise, the EC mechanism likely significantly contributes to the observed gamma-ray flux. According to the 4LAC, the only two repeaters with a CD value of $\sim 1$ are the two ISP BL Lacs: J0509.4+0542 and J1748.6+7005. In several repeaters, the CD value is around 10 or higher (see Table~\ref{tab:res}). This implies that the EC scenario is more favorable as the source of gamma-ray emission in these sources. Even in J1256.1$-$0547, where the CD is relatively low 3.4, nearly all prominent gamma-ray flares between 2008 and 2018 (including our patterns) are very likely of the EC origin \citep{Larionov2020}. Moreover, a detailed SED modeling shows that, for instance, the strongest flares during the fourth repetition of the pattern cannot be explained with a one-zone leptonic model \citep{Hayashida2015,Paliya2016}. The CD based indication of the EC nature of the gamma-ray emission only implies that the seed photons for the IC process are produced outside the emission zone. The broad line region, the dusty torus, and the jet sheath could be the sources of these photons. However, the typical duration of the repeated patterns is $\sim 100$ d in the observer frame. Using proper motions of individual radio-knots of sources in our sample reported by the MOJAVE collaboration \citep{Lister2019}, we estimated that the de-projected distance traveled by an emission feature during an average pattern is between several parsecs and a few tens of parsecs. For instance, in J1256.1$-$0547, the size of the emission region responsible for the pattern is estimated to be 11 pc \citep{Blinov2021}. Such a large size of the emission zone rules out the broad line region and the dusty torus as possible sources of photons for the EC process. Therefore, the jet sheath remains the only feasible option.

Finally, we explore the evolution of the CD value in the sources with repeated patterns. Using optical data from several monitoring projects or transient detection facilities, as described in section~\ref{subsec:opt}, we estimate $CD \approx (\nu F_{\nu})_{\gamma} / (\nu F_{\nu})_{\rm opt}$, where $(F_{\nu})_{\gamma}$ is calculated in the range $2.4 \times 10^{22} \le \nu \le 7.3 \times 10^{25}\, {\rm Hz}$, and $(F_{\nu})_{\rm opt}$ is measured in the range $3.9 \times 10^{14} \le \nu \le 6.3 \times 10^{14}\, {\rm Hz}$. Our CD values may differ significantly from the values reported in  the 4LAC and listed in Table~\ref{tab:res}. This is because, unlike it is done in the 4LAC, we do not fit the synchrotron and inverse Compton peaks. In fact, for most of the LSP sources in our sample, the $(\nu F_{\nu})_{\rm opt}$ value is almost certainly measured at frequencies higher than that of the synchrotron peak. Nevertheless, we are interested in relative changes of the estimated CD with time rather than its absolute value. We plot the obtained CD along with the complete gamma-ray light curves in Figures \ref{fig:0108full} - \ref{fig:0509full}. During several time intervals corresponding to  patterns in different blazars, the CD value demonstrates a clear local maximum. Such intervals include: repetition 2 in J0108.6+0134 (Fig.~\ref{fig:0108full}); repetition 2 in J0457.0-2324 (Fig.~\ref{fig:0457full}); repetitions 2 and 4 in 1256.1-0547 (Fig.~\ref{fig:1256full}); and repetition 2 in 1427.9-4206 (Fig.~\ref{fig:1427full}). In order to demonstrate this unambiguously for the two events in J0505.3+0459 and J0509.4+0542 we computed the mean CD value and its standard error during the second pattern in each of the blazars. We also calculated the same quantities for intervals of equal duration before and after the pattern. These mean CD values with uncertainties are demonstrated in Figures \ref{fig:0505full} and \ref{fig:0509full}. It is clear that in both cases the CD during the pattern is significantly higher compared to its value in adjacent time intervals. This finding is a strong indication that during many of the repeated patterns the Compton dominance value increases. In fact, several gamma-ray flares during the mentioned repeated patterns can be considered as orphan flares, when the optical flux changes insignificantly compared to the corresponding prominent high-energy outburst. Such orphan flares can be explained by the inverse Compton scattering of seed photons provided by the jet sheath during a propagation of an emission feature passing individual overdensities in the sheath \citep{MacDonald2017}.

It is worth mentioning that gravitationally lensed blazars also exhibit repeated patterns of flares in the gamma-ray range \citep{Cheung2014}. In such cases, every prominent flare has to appear twice (or more) in the light curve, with a constant time lag. However, this is not the case for our repeaters. Moreover, the time lags between repetitions of the patterns in our data can reach up to 9 years, which would represent a record long delay among known lenses. Such a delay would imply a very large lensing mass and a large angular separation between the lensed images \citep{Munoz2022}. In this scenario, these blazars would have been identified as lensed sources a long time ago.

We also note that two blazars among our repeaters have been found to posses a significant periodic signal in their gamma-ray light curves. \cite{Penil2022} reported on the presence of periodicity in J0457.0-2324. However, this period appears to be unrelated to the repeated pattern of the source. \cite{Wang2022} found a $\sim 3.1$ year period in the gamma-ray data of J1048.4+7143. This detection is based on the three major flares of this source that are clearly visible in Fig.~\ref{fig:1048full}. Our patterns are detected in the second halves of the first two of these major flares. \cite{Kun2022} analysed multi-frequency data of J1048.4+7143 and suggested an ongoing merging of a supermassive binary black hole in this source. This merging is presumably causing the jet precession, which is reflected in the gamma-ray light curve. Potentially, this scenario does not contradict the possibility of the spine-sheath structure of the jet in this source.

The properties of the repeated patterns and the corresponding blazars discussed in this section favour the spine-sheath structured jet scenario for the interpretation of the observed events. 
Nevertheless, we cannot fully dismiss the alternative scenario. It is possible that in certain blazars, such as J1748.6+7005, patterns of flares might be generated by the collision of moving features with consecutive recollimation shocks. Additionally, as shown by \cite{Hervet2017}, these two models could be interconnected. The existence of recollimation shocks in the jet spine is dependent upon the relative power of the sheath and spine.

\section{Discussion and conclusions} \label{sec:disc}

The possibility of velocity stratification in relativistic jets has been discussed in the literature for a long time. In certain models of jet formation, a sheath around the jet spine with distinct kinematic properties is produced naturally \citep{Sol1989,Aloy2000,McKinney2006}. On the other hand, the "limb brightening" observed with VLBI in some AGN can be considered as observational evidence that such spine-sheath jets indeed exist \citep{Attridge1999,Giroletti2004,Bruni2021}. In this work, for the first time we provide evidence of spine-sheath structure in the inner jets of a large fraction ($\gtrsim10$\%) of blazars, based on gamma-ray light curves. This finding has several important implications that we discuss below.

 As theoretically predicted \citep{Tavecchio2014,Tavecchio2015}, an interaction between the spine and the sheath in relativistic jets can result in a high neutrino luminosity. This model is one of the possible scenarios where protons accelerated in the jet interact with a dense photon field external to the jet. Alternatively, this photon field can be produced by the BLR \citep{Murase2014,Padovani2019} or by radiatively inefficient accretion flows \citep{Righi2019}. The presence of an external photon field can significantly reduce the power requirements for neutrino production in AGN jets and is favored by various models \citep{Keivani2018,Petropoulou2020}. Our work provides compelling evidence for the spine-sheath scenario as the explanation for neutrino production in jets, as we have discovered repeated patterns of gamma-ray flares that are likely associated with neutrino events at the $2.8\sigma$ level. This finding is further supported by the recent discovery of the limb brightening in TXS 0506+056 \citep{Ros2020}, where we also found a repeated pattern that was ongoing during the moment of arrival of IC170922A. The latter neutrino is the most confidently associated event with a flaring blazar to date \citep{Aartsen2018a}. We emphasize that timing information is very important for association of neutrinos and blazars. In our case only two events provide the $2.8 \sigma$ significance level. Increasing the number of detected repeaters by a factor of a few in the near future could potentially increase the confidence of the result up to $5 \sigma$, which would be impossible without the temporal information \citep{Liodakis2022}.

 The nature of gamma-ray emission in blazars is still under debate \cite[e.g.,][and references therein]{Cerruti2020}. The repeated patterns we report here place important constraints on the mechanism and location of the emission zone (see section \ref{sec:mech}). Our findings suggest that a considerable portion of blazars, amounting to $\gtrsim10$\%, generate EC gamma-ray flares over a considerable distance in their jet, ranging from several parsecs to a few tens of parsecs. Furthermore, the events discussed in this work can explain a significant fraction of orphan gamma-ray flares, which may constitute up to $\sim 20$\% of all high energy flares in blazars \citep{Liodakis2019,Jaeger2022}.

The repeated patterns of gamma-ray flares can be used to study the inner jet kinematics in regions that were previously unavailable for VLBI before the Event Horizon Telescope due to opacity at low radio frequencies. The time scale contraction or expansion factors that we find juxtaposing repeating patterns of the same source are inversely proportional to the ratio of Doppler factors of the corresponding moving emission features (see section \ref{sec:meth}). Therefore, by associating patterns with individual radio knots ejections in VLBI radio maps, as has been done in \cite{Jorstad2013} or \cite{Blinov2021}, one could study how the Doppler factors of individual components change in the acceleration and collimation zone.

Moreover, the repeated patterns provide us a way to perform a 1-D tomography of the parsec-scale jet sheath and observe its evolution, enabling testing of jet launching and propagation models. As recently demonstrated by \cite{Boccardi2021}, jets in High Excitation Radio Galaxies (HERG) are collimated on larger scales compared to Low Excitation Radio Galaxies (LERG). They explain this difference by a presence of powerful collimated disk winds (or sheaths) in HERGs that help their jets to remain confined at longer distances from the central engine \citep{Globus2016}. It is likely that these winds exist in both LERGs and HERGs, but drastic differences in the accretion regime of these classes regulate their characteristics. In HERGs these winds have orders of magnitude larger radius and carry more power with respect to LERGs. Similar results were found by \cite{Potter2015} for FSRQs and BL Lacs, which constitute the beamed counterparts of HERGs and LERGs. Our results demonstrate that most of the sheaths 8 \cite[or 9 if TXS 0506+056 is considered as an FSRQ following][]{Padovani2019} out of 10 are found in FSRQs. However, this may still be an accidental outcome of the highly biased parent sample. Further increasing the number of detected repeaters could help verify whether FSRQs indeed tend to have the spine-sheath structure more frequently than BL Lac sources and study what parameters regulate properties of these sheaths.


As mentioned in section \ref{sec:neutr}, some of the visually identified patterns do not follow the assumed dependence the flares amplitude on the time scale $\Delta t$ changes for repetitions detected in the same source. For instance, the second pattern repetition in J0509.4+0542 lasted longer in time compared to the first. Contrary to the expected $\propto \Delta t^{-(3+\alpha)}$ or $\propto \Delta t^{-(4+2\alpha)}$ behaviour (see section \ref{sec:meth}) of amplitudes, we found them significantly larger in the second repetition with respect to the first. A similar discrepancy is observed for other patterns (e.g., in J0457.0-2324). Our overly simplistic model of the repeated patterns may be the reason behind the observed inconsistency. According to \cite{Ghisellini2005}, the observed monochromatic intensity in the case of the EC mechanism is:
\begin{equation}
    I(\nu) = I^\prime(\nu^\prime) \delta^{4+2\alpha}_{\rm s,sh} \delta^{3+\alpha}_{\rm sh},
\end{equation}
where $I^\prime(\nu^\prime)$ is the monochromatic intrinsic intensity produced by the spine, $\delta_{\rm s,sh}$ is the Doppler factor of the spine in the reference of the sheath, and $\delta_{\rm sh}$ is the Doppler factor of the sheath in the observer's frame. In section \ref{sec:meth}, we assumed a non-relativistic sheath scenario $\delta_{\rm sh} = 1$, simplifying this equation. However, this may not be the case in reality, where the sheath can be mildly relativistic. The sheath can accelerate or decelerate with time given the long time intervals between observed patterns. Moreover,  individual overdensities in the sheath (which we believe are responsible for individual flares in the patterns) can evolve with time, as well as $I^\prime(\nu^\prime)$, which will further complicate the observed dependencies. Future attempts to search for repeated patterns should consider a more realistic model, which may result in a much larger sample of detected events.

In summary, we conducted a systematic search for repeated patterns of flares in gamma-ray light curves of the brightest blazars. We found 8 new sources that exhibit such patterns, in addition to the two previously known sources. Of the ten sources with repeated patterns, half can be potentially associated with neutrino events discussed in the literature. An MC simulation performed with a statistically well defined sample of track-like neutrinos from the ICECAT-1 catalogue suggests their link to the repeated patterns at the $2.8\sigma$ significance level.

\section*{Acknowledgments}

We thank I. Liodakis for useful comments. This project is supported by the Russian Science Foundation grant 23-22-00121. D.B. acknowledges support from the European Research Council (ERC) under the European Union Horizon 2020 research and innovation program under the grant agreement No 771282. This work is based on Fermi data, obtained from Fermi Science Support Center, provided by NASA’s Goddard Space Flight Center
(GSFC). We acknowledge the hard work by the Fermi-LAT Collaboration that provided the community with unprecedented quality data and made Fermi Tools so readily available. This publication makes use of data products from the Wide-field Infrared Survey Explorer, which is a joint project of the University of California, Los Angeles, and the Jet Propulsion Laboratory/California Institute of Technology, funded by the National Aeronautics and Space Administration. This publication makes use of data products from the Two Micron All Sky Survey, which is a joint project of the University of Massachusetts and the Infrared Processing and Analysis Center/California Institute of Technology, funded by the National Aeronautics and Space Administration and the National Science Foundation.

\section*{Data Availability}

The gamma-ray light curves analysed in this article are available in Harvard
Dataverse via \url{https://doi.org/10.7910/DVN/JMJVCH}. Additionally, the Fermi LAT Light Curve Repository contains gamma-ray light curves for the 100 brightest blazars with 3 and 7 d integration times. Optical and neutrino data are publicly available on the web pages of the corresponding facilities.



\bibliographystyle{mnras}
\bibliography{patterns}




\appendix

\section{}

In this section, we present the entire {\em Fermi}-LAT gamma-ray light curves of the blazars with repeating patterns found.

\begin{figure*}
	\includegraphics[width=1.0\textwidth]{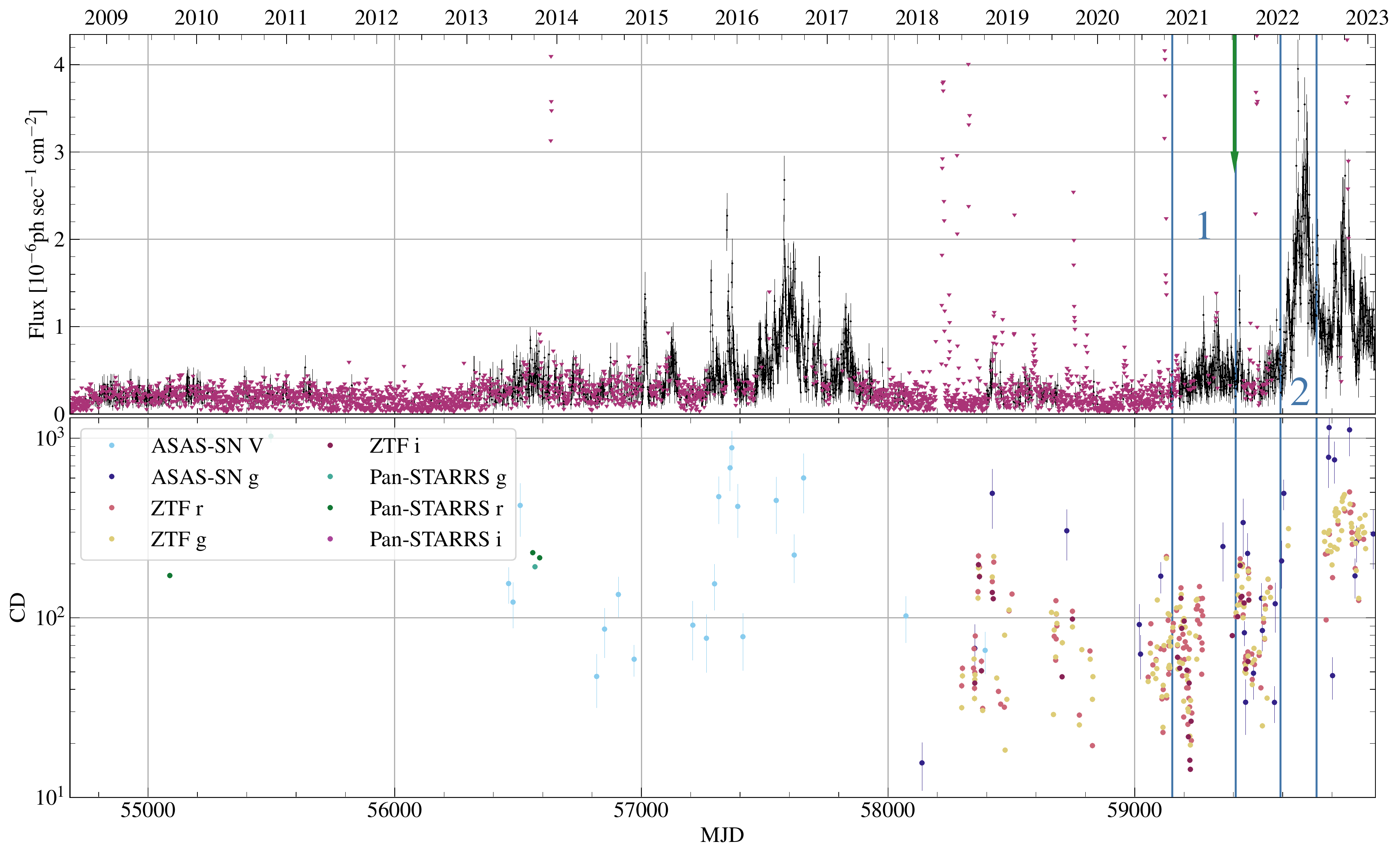}
    \caption{(Top panel) The entire light curve of 4FGL J0108.6+0134. The green arrow indicates the moment of arrival of GVD210710CA. (Bottom panel) Evolution of the Compton dominance.}
    \label{fig:0108full}
\end{figure*}

\begin{figure*}
	\includegraphics[width=1.0\textwidth]{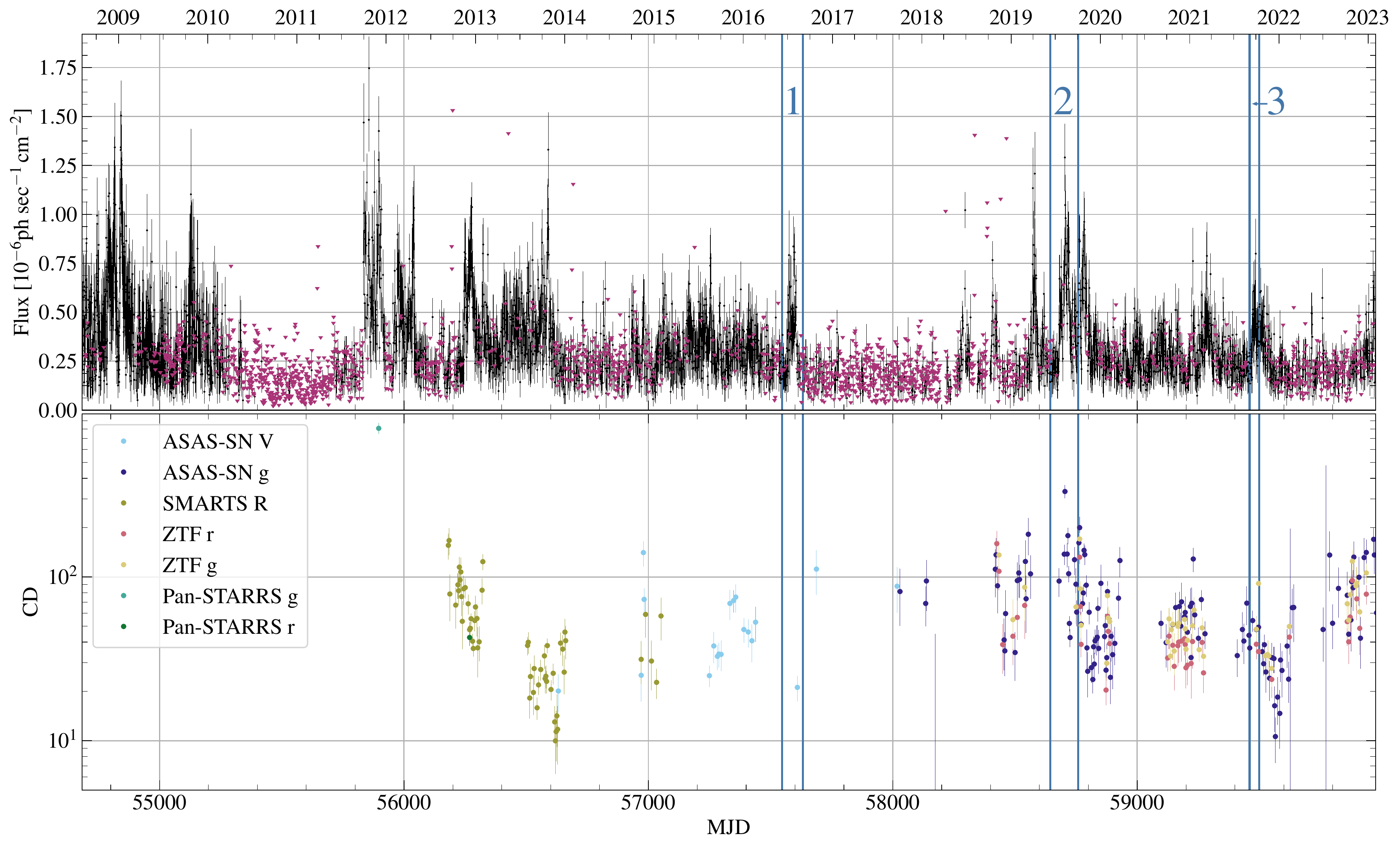}
    \caption{(Top panel) The entire light curve of 4FGL J0457.0-2324. (Bottom panel) Evolution of the Compton dominance.}
    \label{fig:0457full}
\end{figure*}

\begin{figure*}
	\includegraphics[width=1.0\textwidth]{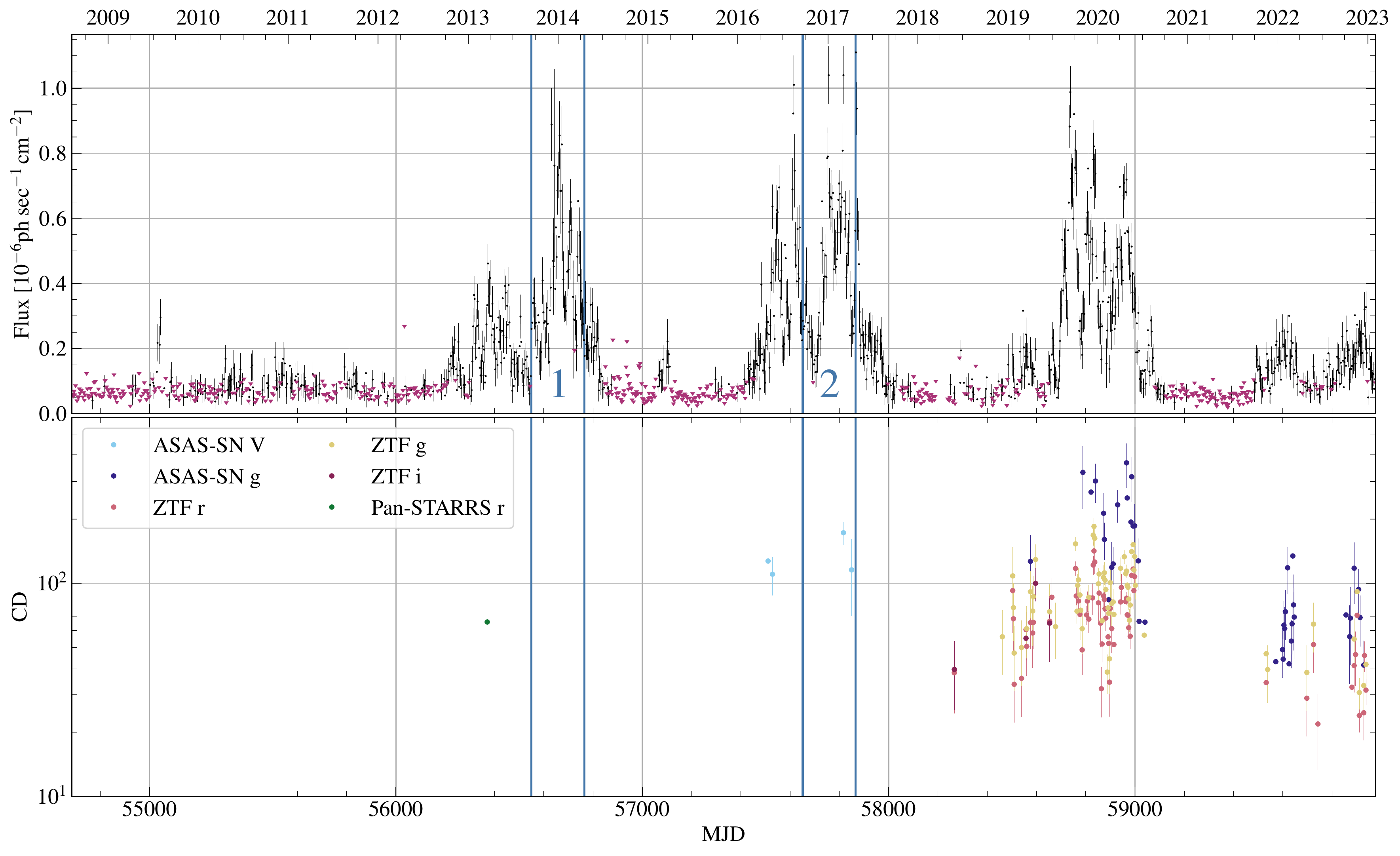}
    \caption{(Top panel) The entire light curve of 4FGL J1048.4+7143. (Bottom panel) Evolution of the Compton dominance.}
    \label{fig:1048full}
\end{figure*}

\begin{figure*}
	\includegraphics[width=1.0\textwidth]{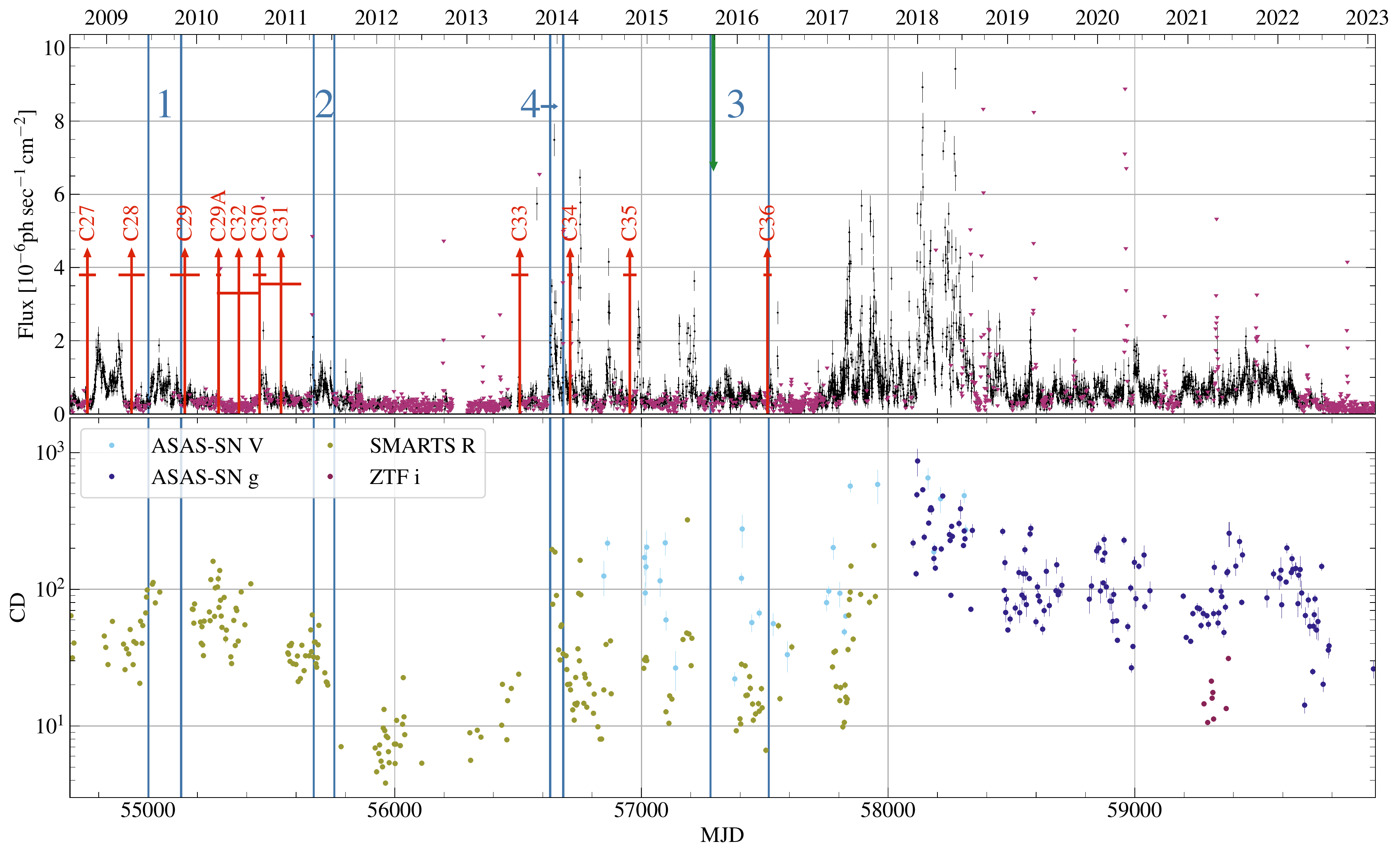}
    \caption{(Top panel) The entire light curve of 4FGL J1256.1-0547. The green arrow indicates the moment of the IC150926A neutrino event at MJD=57291.9. The red arrows indicate the moments of new radio knots ejection originally presented by \protect\cite{Jorstad2017} and analysed in \protect\cite{Blinov2021}. (Bottom panel) Evolution of the Compton dominance.}
    \label{fig:1256full}
\end{figure*}

\begin{figure*}
	\includegraphics[width=1.0\textwidth]{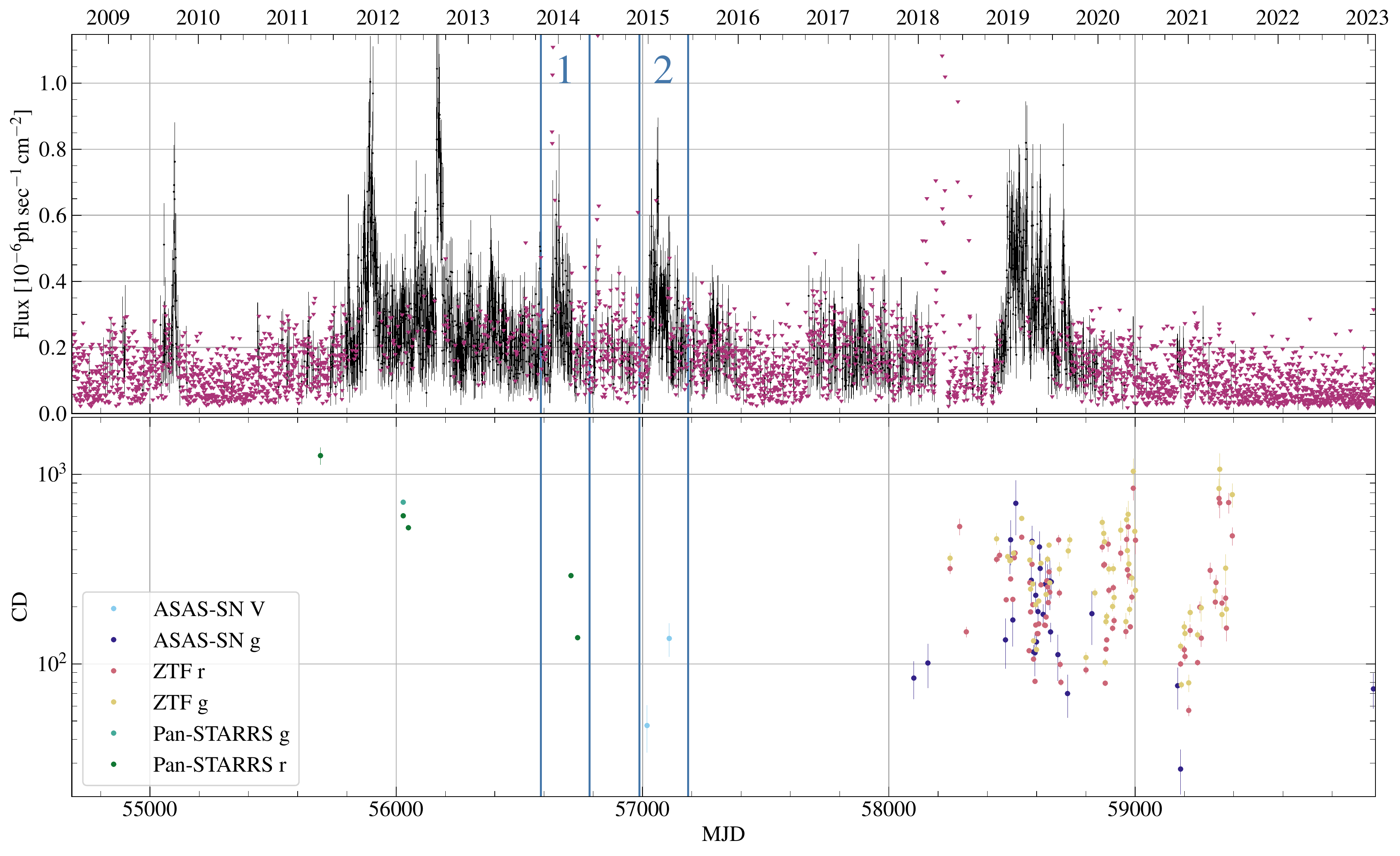}
    \caption{(Top panel) The entire light curve for 4FGL J1345.5+4453. (Bottom panel) Evolution of the Compton dominance.}
    \label{fig:1345full}
\end{figure*}

\begin{figure*}
	\includegraphics[width=1.0\textwidth]{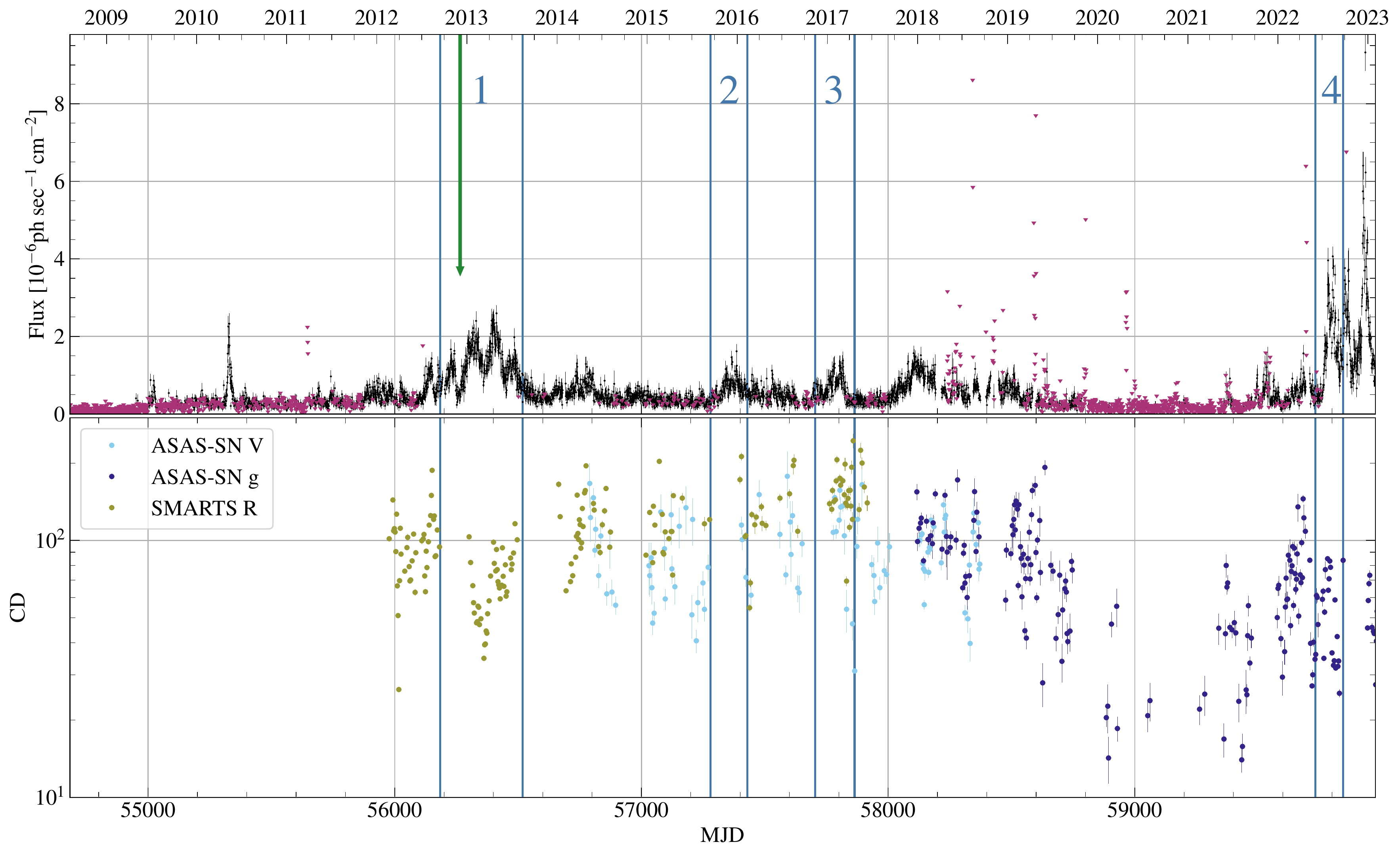}
    \caption{(Top panel) The entire light curve of 4FGL 1427.9-4206. The green arrow indicates the moment of arrival of IC-35 ("Big Bird"). (Bottom panel) Evolution of the Compton dominance.}
    \label{fig:1427full}
\end{figure*}

\begin{figure*}
	\includegraphics[width=1.0\textwidth]{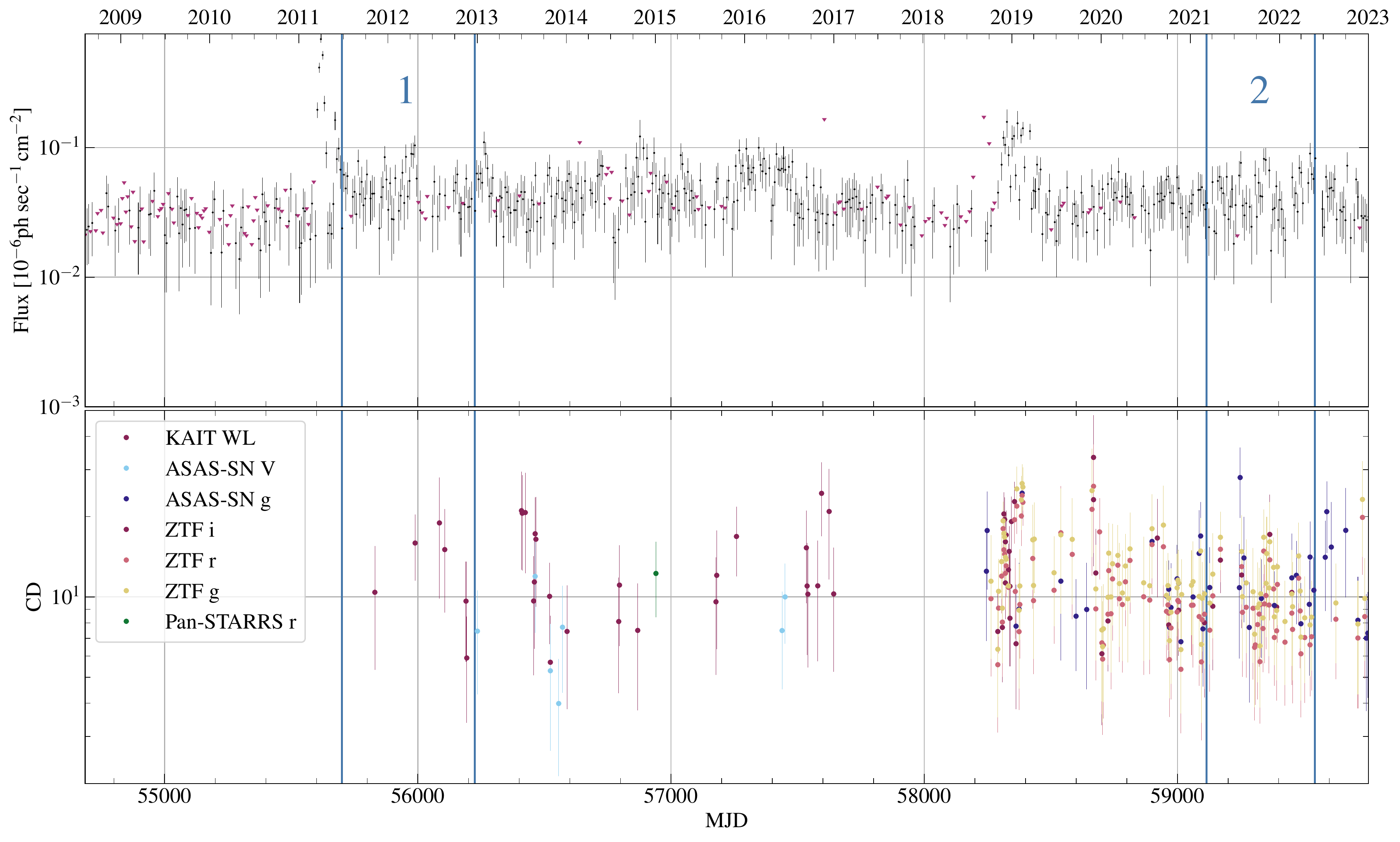}
    \caption{(Top panel) The entire light curve of 4FGL 1748.6+7005. (Bottom panel) Evolution of the Compton dominance.}
    \label{fig:1748full}
\end{figure*}

\begin{figure*}
	\includegraphics[width=1.0\textwidth]{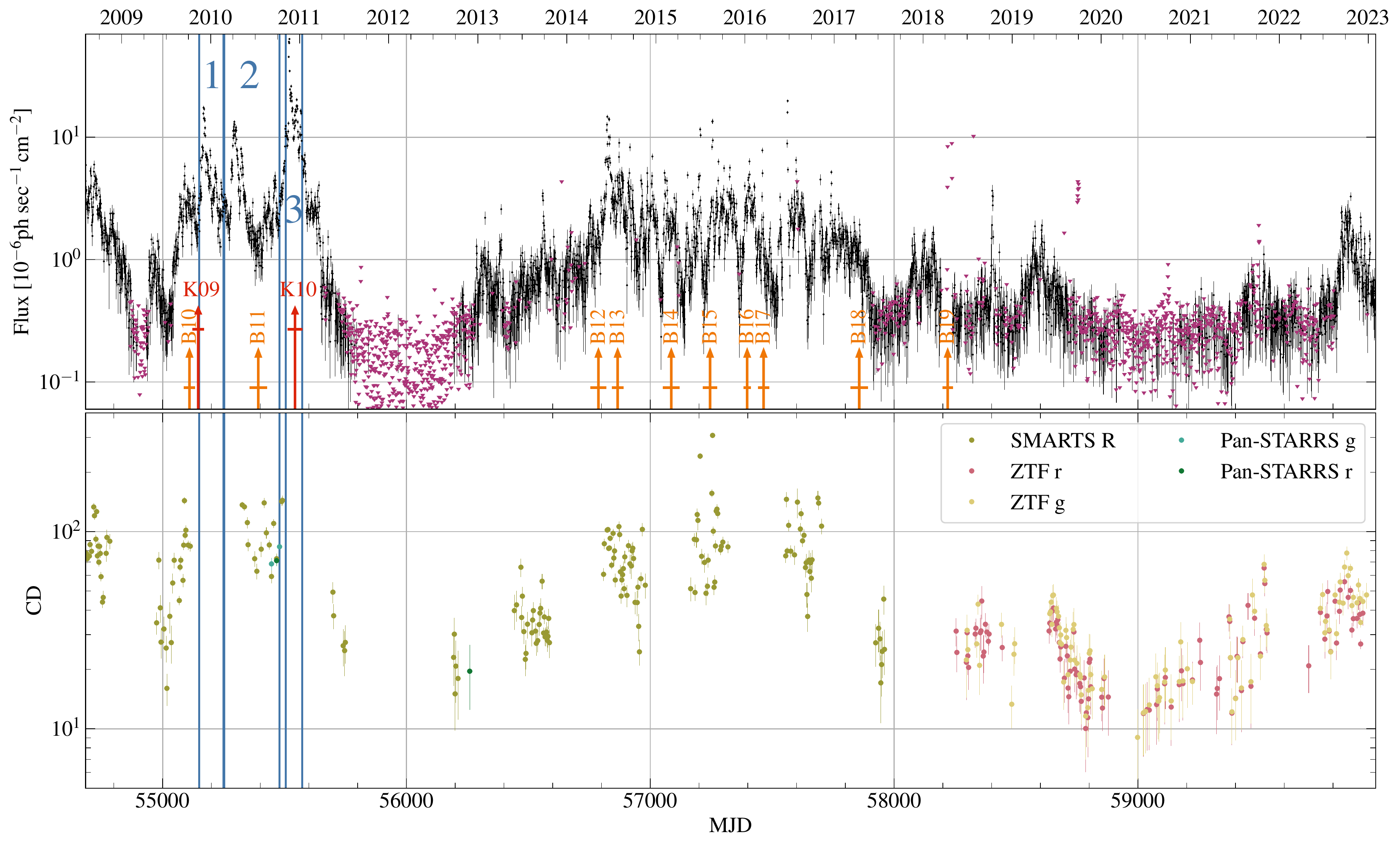}
    \caption{(Top panel) The entire light curve of 4FGL 2253.9+1609. The red arrows indicate the moments of new radio knots ejection from \protect\cite{Jorstad2013}. The orange arrows indicate the moments of new radio knots ejection from \protect\cite{Weaver2022}. (Bottom panel) Evolution of the Compton dominance.}
    \label{fig:2253full}
\end{figure*}

\begin{figure*}
	\includegraphics[width=1.0\textwidth]{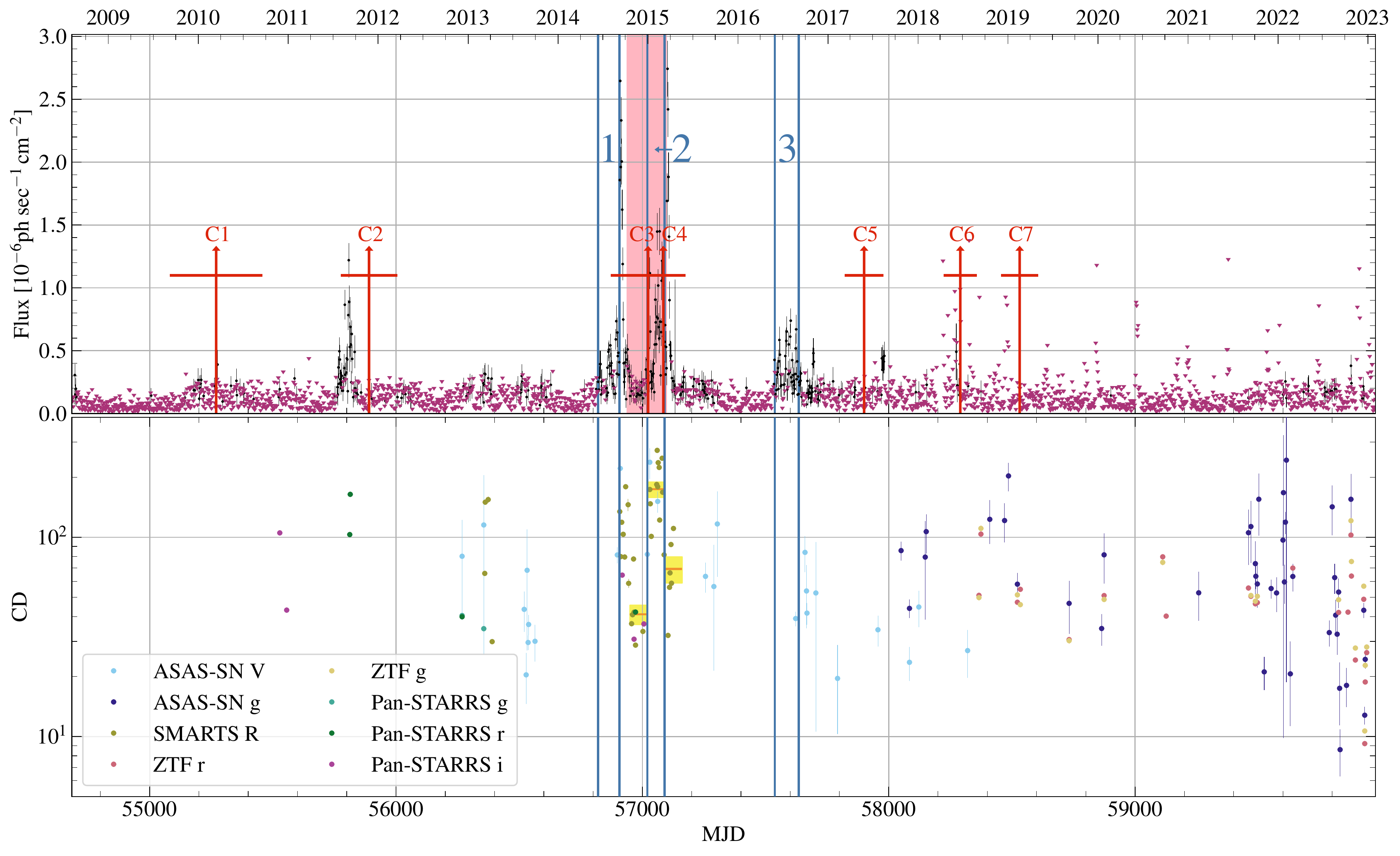}
    \caption{(Top panel) The entire light curve of 4FGL J0505.3+0459. The pink region marks the $\sim 3.7\sigma$ significance neutrino excess interval from MJD=56937.8 to MJD=57096.2 reported in \protect\cite{Aartsen2018b}. The red arrows indicate the moments of new radio knots ejection from \protect\cite{Sumida2022}. (Bottom panel) Evolution of the Compton dominance. The orange horizontal lines show the mean CD value during the second pattern interval and during time intervals of the same duration before and after it. The yellow areas of show the standard error of the mean CD value during these intervals.}
    \label{fig:0505full}
\end{figure*}

\begin{figure*}
    \includegraphics[width=1.0\textwidth]{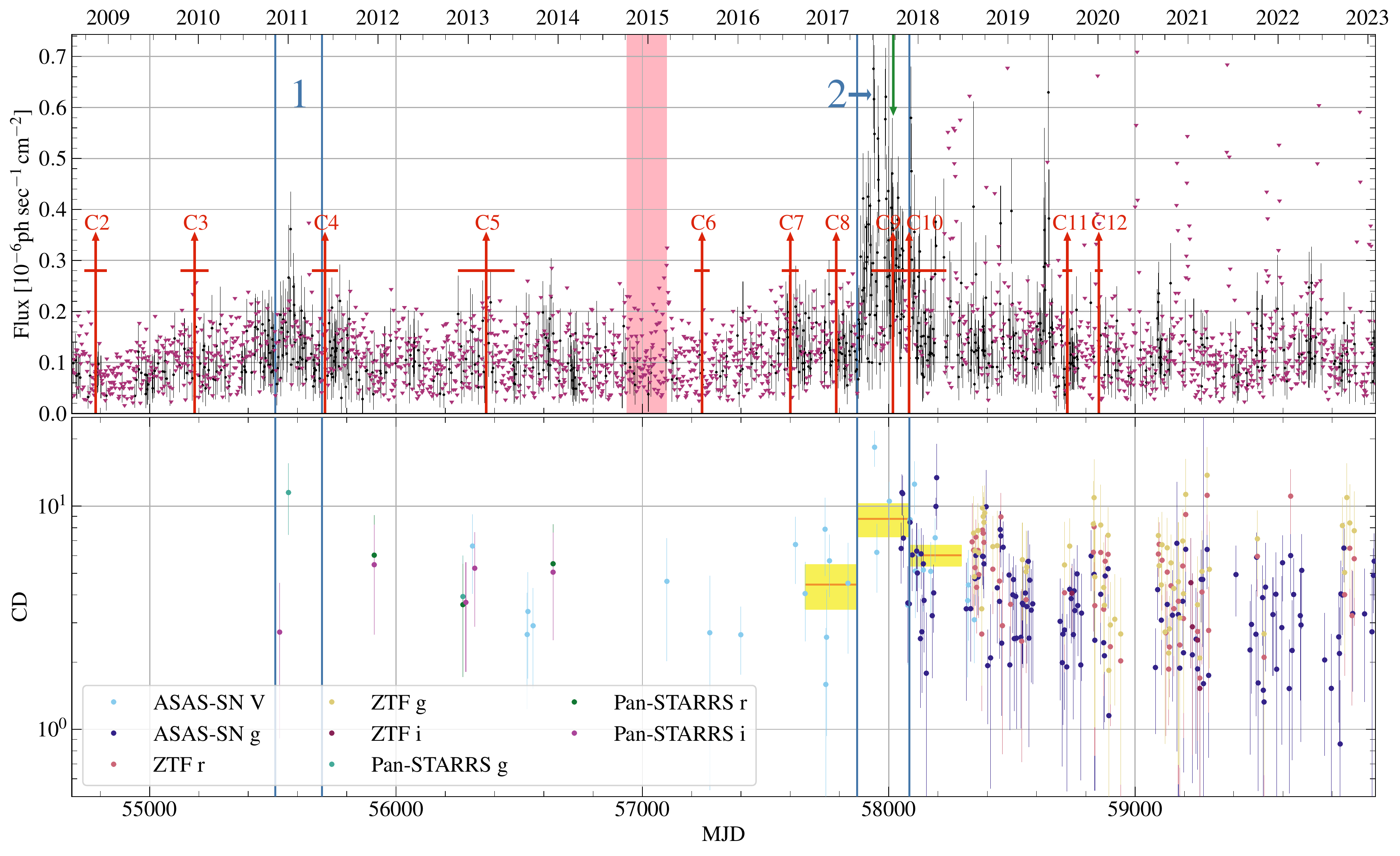}
    \caption{(Top panel) The entire light curve of 4FGL J0509.4+0542. The green arrow indicates the moment of arrival of IceCube-170922A. The pink region marks the $\sim 3.7\sigma$ significance neutrino excess interval from MJD=56937.8 to MJD=57096.2 reported in \protect\cite{Aartsen2018b}. The red arrows indicate the moments of new radio knots ejection from \protect\cite{Sumida2022}. (Bottom panel) Evolution of the Compton dominance. The orange horizontal lines show the mean CD value during the second pattern interval and during time intervals of the same duration before and after it. The yellow areas of show the standard error of the mean CD value during these intervals.}
    \label{fig:0509full}
\end{figure*}


\bsp	
\label{lastpage}
\end{document}